\documentclass[a4paper,10pt,fleqn]{article}

\usepackage{blindtext}%

\usepackage[numbers]{natbib}%
\usepackage[bottom=3cm,top=2cm,left=1.5cm,right=1.5cm]{geometry}%
\usepackage{framed,multirow,url,setspace,graphicx,bm,xcolor,siunitx,nicefrac,multicol}
\usepackage{amsmath,amssymb,amsthm}%
\usepackage{stmaryrd,textcomp,multido,colortbl,ifthen}%

\usepackage[hang]{subfigure}

\usepackage[english]{babel}
\usepackage[utf8]{inputenc}

\usepackage{notation}%
\usepackage{mathoperators}%
\usepackage{reconstructnotation}%

\usepackage{hyperref}

\newcommand{\reffig}[1]{figure~\ref{fig:#1}}%
\newcommand{\refFig}[1]{Figure~\ref{fig:#1}}%
\newcommand{\reftab}[1]{table~\ref{tab:#1}}%
\newcommand{\refTab}[1]{Table~\ref{tab:#1}}%
\newcommand{\refEqn}[1]{Equation~\ref{eqn:#1}}%
\newcommand{\refeqs}[1]{eqs.~\ref{eqn:#1}}%
\newcommand{\refeqno}[1]{\ref{eqn:#1}}%
\makeatletter
\newcommand{\refeqn}{\@ifstar{\@ifstar\@@refeqn\@refeqn}{\@@@refeqn}}%
\def\@refeqn#1{(\ref{eqn:#1})}%
\def\@@refeqn#1{eqs.~(\ref{eqn:#1})}%
\def\@@@refeqn#1{eq.~(\ref{eqn:#1})}%
\makeatother

\newcommand{\refeqnsave}[1]{eq.~(\ref{eqn:#1})}%

\colorlet{shadecolor}{black!12}%
\newtheorem{notetheorem}{Note}[section]
\newenvironment{note}%
{\colorlet{shadecolor}{black!12}\begin{shaded}\begin{notetheorem}}%
{\end{notetheorem}\end{shaded}}%
\newcommand{\refnote}[1]{note~\ref{note:#1}}%
\newcommand{\refNote}[1]{Note~\ref{note:#1}}%

\newtheorem{remarktheorem}{Remark}[section]%
\newenvironment{remark}%
{\colorlet{shadecolor}{black!12}\begin{shaded}\begin{remarktheorem}}%
{\end{remarktheorem}\end{shaded}}%

\newtheorem*{impltheo}{Implementation detail}
\newenvironment{impldetail}%
{\colorlet{shadecolor}{black!12}\begin{shaded}\begin{impltheo}}%
{\end{impltheo}\end{shaded}}

\def\updatefigure#1{}

\updatefigure{346}%
\updatefigure{347}%
\updatefigure{348}%
\updatefigure{349}%
\updatefigure{350}%
\updatefigure{351}%
\updatefigure{352}%
\updatefigure{353}%
\updatefigure{354}%
\updatefigure{359}%

\definecolor{newcolor}{rgb}{.8,.349,.1}

\def\worktitle{Face-based Volume-of-Fluid interface positioning in arbitrary polyhedra}%

\providecommand{\keywords}[1]{\textbf{\textit{Keywords---}} #1}%

\allowdisplaybreaks%

\begin{document}
\thispagestyle{empty}%

\title{\worktitle}%
\author{Johannes Kromer\href{https://orcid.org/0000-0002-6147-0159}{\includegraphics[height=10pt]{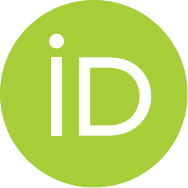}} and Dieter Bothe\textsuperscript{$\dagger$}\href{https://orcid.org/0000-0003-1691-8257}{\includegraphics[height=10pt]{orcid_logo}}}%
\date{}%
\maketitle
\begin{center}
Mathematical Modeling and Analysis, Technische Universität Darmstadt\\ Alarich-Weiss-Strasse 10, 64287 Darmstadt, Germany\\%
\textsuperscript{$\dagger$}Email for correspondence: \href{mailto:bothe@mma.tu-darmstadt.de?subject=Face-based\%20Volume-of-Fluid\%20interface\%20positioning\%20in\%20arbitrary\%20polyhedra}{bothe@mma.tu-darmstadt.de}%
\end{center}

\begin{abstract}%
We introduce a fast and robust algorithm for finding a plane $\plicplane$ with given normal $\plicnormal$, which truncates an arbitrary polyhedron $\polyhedron*$ such that the remaining sub-polyhedron admits a given volume $\refvof\abs{\polyhedron*}$. In the literature, this is commonly referred to as \textbf{Volume-of-Fluid (VoF) interface positioning problem}. The novelty of our work is twofold: %
firstly, by recursive application of the \textsc{Gaussian} divergence theorem, the volume of a truncated polyhedron can be computed at high efficiency, based on summation over quantities associated to the faces of the polyhedron. One obtains a very convenient piecewise parametrization (within so-called \textit{brackets}) in terms of the signed distance $\signdist$ to the plane $\plicplane$. %
As an implication, one can restrain from the costly necessity to establish topological connectivity, rendering the present approach most suitable for the application to unstructured computational meshes. 
Secondly, in the vicinity of the truncation position $\signdist$, the volume can be expressed exactly, i.e.\ in terms of a cubic polynomial of the normal distance to the PLIC plane. The local knowledge of derivatives enables to construct a root-finding algorithm that pairs bracketing and higher-order approximation. %
The performance is assessed by conducting an extensive set of numerical experiments, considering convex and non-convex polyhedra of genus (i.e., number of holes) zero and one in combination with carefully selected volume fractions $\refvof$ (including $\refvof\approx0$ and $\refvof\approx1$) and normal orientations $\plicnormal$. For all configurations we obtain a significant reduction of the number of (computationally costly) truncations required for the positioning: on average, our algorithm requires between one and two polyhedron truncations to find the position of the plane $\plicplane$, outperforming existing methods. %
\end{abstract}%



%
%
\section{Introduction}\label{sec:introduction}%
In geometric Volume-of-Fluid (VoF) methods, the approximative reconstruction of the interface from volume fractions constitutes a central element, in terms of both accuracy of the method and consumption of computational ressources. %
%
%
In its most extensive scope, the reconstruction of an approximative interface includes the computation of the normal field from the volume fractions, followed by positioning the planar interface patches such that they match the respective volume fraction in each cell. The present work focusses on the second component, assuming the discrete field of approximative normals to be given. %
As we shall see below, positioning the interface translates to finding the unique root of a scalar monotonous function, namely the volume of the truncated computational cell, which is parametrized by the signed distance $\signdist$. However, the truncated volume is only known implicitly, in the sense that its evaluation at some $\signdist$ requires a (computationally costly) truncation of the polyhedron. Therefore, there are two conceptually independent components: %
\begin{description}%
%
%
\item[Volume computation of a truncated polyhedron:] %
The vast majority of literature contributions, in one form or another, resorts to schemes involving truncation, (i) establishing topological connectivity and (ii) volume computation. The combinatorial complexity of task (i) depends on the topological properties of the original polyhedron as well as on the possible truncation outcomes. In the present context, computing the volume of an arbitrary polyhedron by application of the \textsc{Gaussian} divergence theorem can hardly be outplayed in terms of efficiency; cf.~\citet{JCP_2019_ncaa}. While explicit formulae can be given for primitive polyhedra (e.g., \citet{JCP_2006_arfr,JCP_1999_vofi}), tetrahedral decomposition is applied in general\footnote{Figure 6 in \citet{JCP_2007_mmir} contains a nice illustration of the implied complexity.}, being both costly and limited to topologically simple or convex polyhedra in most cases (e.g., \citet{JCP_2018_aiir} and tables~3 and 4 in \citet{JCP_2019_ncaa}). %
\item[Root-finding:] The choice of an appropriate method depends, among others, on the available information: while the secant method gets along with function evaluations, \textsc{Newtons} method or consecutive cubic spline interpolation (e.g., \citet{XXX_2020_ivof}) additionally require derivatives. %
For a thorough overview, the reader is referred to the books of \citet{numerical_analysis_stoer} and \citet{numerical_analysis_atkinson}. %
\end{description}%
The fact that the interface needs to be reconstructed within each computational cell and time step\footnote{\citet{JNA_1968_otca} introduces a directionally split geometric transport, requiring three reconstructions per time step.} thus highlights the necessity for both general applicability (in terms of cell topology) and computational efficiency. %
%
%
The application of the \textsc{Gaussian} divergence theorem allows to address both of the aforementioned issues. As stated above, this is a common concept in the literature. However, the present approach introduces the exploitation of a previously neglected scope for design: pairing recursive application of a reduction in dimension via the \textsc{Gaussian} divergence theorem with proper translations of the coordinate system avoids the costly and, thus, disadvantageous necessity to extract topological information from the truncated polyhedron. While this property considerably reduces the complexity for convex polyhedra already, its full potential develops for non-convex polyhedra or those with a large number of faces. Furthermore, our method produces the derivatives with respect to $\signdist$ of the volume function at negligible additional costs, enabling the application of a root-finding algortihm based on locally quadratic approximation in combination with implicit bracketing\footnote{The term \textit{bracketing} refers to finding the bracket containing the sought position $\signdistref$.}.%
%
%

The remainder of this paper is organized as follows: after reviewing the relevant literature in subsection~\ref{subsec:literature_review}, the formal notation and data structure are summarized in subsection~\ref{subsec:notation}. Section~\ref{sec:mathematical_formulation} provides the mathematical details of the volume computation, along with a novel rigorous analytical treatment of the parametrized volume function. %
Since this paper introduces a numerical method meant to be integrated into existing codes, section~\ref{sec:implementation} offers a detailed flowchart, which is further detailed throughout this work. %
In section~\ref{sec:numerical_results}, we present numerical results a set of convex and non-convex polyhedra. %
%
%
\subsubsection*{Graphics}%
The figures and flowcharts in this manuscript are produced using the versatile and powerful library \href{https://ctan.org/topic/pstricks}{\texttt{pstricks}}. For implementation details as well as an instructive collection of examples, the reader is referred to the book of \citet{pstricks_2008} and \url{https://tug.org/PSTricks/main.cgi}, respectively. %
%
%
\subsection{Literature review}\label{subsec:literature_review}%
\begin{figure}[htbp]
\null\hfill%
\includegraphics[page=1]{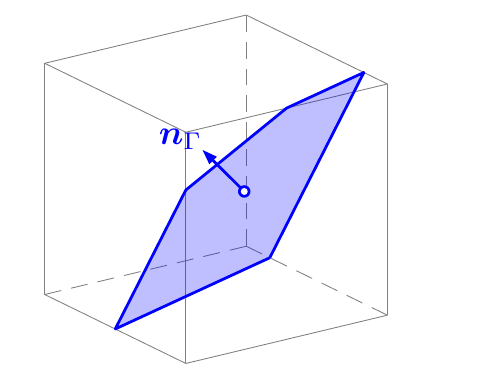}%
\hfill%
\includegraphics[page=2]{diot_rotation}%
\hfill%
\includegraphics[page=3]{diot_rotation}%
\hfill\null%
\caption{Rotation of cuboid s.t.\ $\plicnormal\to\ve_z$; center: planes at $\signdist_{k}$/$\signdist_{k+1}$ (red, passing through cell vertices \textcolor{red}{$\blacksquare$}) bracketing the sought plane $\plicplane$ (blue); right: prismatoid (obtained from two consecutive planes truncating the cuboid) whose volume is computed analytically; cf.~\protect\citet{JCP_2016_airm}.}%
\label{fig:diot_rotation_bracketing}
\end{figure}

\begin{figure}[htbp]%
\null\hfill%
\includegraphics[page=1]{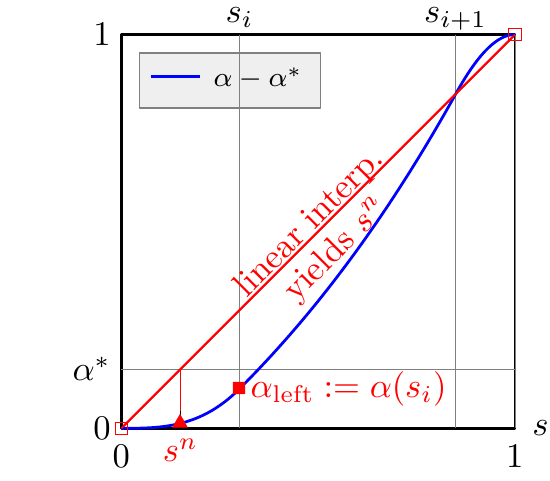}%
\hfill%
\includegraphics[page=2]{cibrave_step}%
\hfill%
\includegraphics[page=3]{cibrave_step}%
\hfill\null%
\caption{CIBRAVE method proposed by \protect\citet{JCP_2016_anvc}: the first iteration $\signdist^{n}$ (left, $\textcolor{red}{\blacktriangle}$) lies not within the target bracket $\refbracket=[\projvertpos{i},\projvertpos{i+1}]$. Due to $\polyvof\fof{\signdist^n}<\refvof$, $\polyvof_{\mathrm{left}}$ is updated, such that a linear interpolation yields $\signdist^{n+1}$ (center, $\textcolor{red}{\blacktriangle}$). Since $\polyvof\fof{\projvertpos{i}}\leq\polyvof\fof{\signdist^{n+1}}\leq\polyvof\fof{\projvertpos{i+1}}$, the polyhedron is truncated at $\projvertpos{i}$. Finally, due to $\polyvof\fof{\projvertpos{i}}<\refvof<\polyvof\fof{\projvertpos{i+1}}$, a cubic spline interpolation yields the sought $\signdistref$ (right, $\textcolor{red}{\bullet}$), resulting in two polyhedron truncations.}%
\label{fig:cibrave_step_illustration}%
\end{figure}

\begin{figure}[htbp]%
\null\hfill%
\includegraphics[page=1]{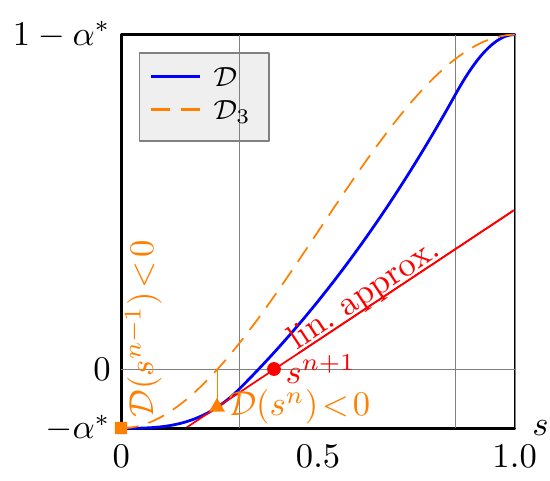}%
\hfill%
\includegraphics[page=2]{maric_consecutive_step}%
\hfill%
\includegraphics[page=3]{maric_consecutive_step}%
\hfill\null%
\caption{CCS method proposed by \protect\citet{XXX_2020_ivof}: computation of next iteration ($\signdist^{n+1}$, $\textcolor{red}{\bullet}$) for coinciding (left) and different (center) sign of deviation $\vofdeviation$ at current ($\signdist^{n}$, $\textcolor{orange}{\blacktriangle}$) and previous iteration ($\signdist^{n-1}$, $\textcolor{orange}{\blacksquare}$). If the interval ${[\signdist_{\mathrm{left}},\signdist_{\mathrm{right}}]}$ is confined by a bracket, the spline interpolates $\vofdeviation$ exactly (right).}%
\label{fig:ccs_step_illustration}%
\end{figure}

\begin{note}[Problem formulation]\label{note:problem_formulation}%
In a VoF context, the phases are encoded by the discrete volume fraction field $\alpha$, from which, in geometric methods, at some point the numerical interface has to be reconstructed. The aforementioned reconstruction involves (i) the computation of the normal field from the volume fractions, followed by (ii) a piece-wise (i.e.\ cell-based) positioning of the disconnected numerical interface patches, ensuring that the volume fractions $\polyvof$ in the respective cells are conserved to a prescribed tolerance. The present work focusses on the latter step. %
Hence, the problem under consideration, also referred to as \textbf{volume-matching} (\citet{JCP_2016_airm}) or \textbf{volume conservation enforcement (VCE)} (\citet{JCP_2019_ncaa}) in the literature, can be cast as follows: given %
\begin{itemize}
\item a polyhedron $\polyhedron*$,%
\item a unit normal $\plicnormal$ and %
\item a volume $0\leq\refvol\leq\volume{\polyhedron*}$ or, alternatively, a volume fraction $0\leq\refvof\leq1$ with $\refvol=\refvof\volume{\polyhedron*}$,%
\end{itemize}
find some $\signdistref\in\setR$ such that the volume of the intersection of the polyhedron $\polyhedron*$ with the negative halfspace of the PLIC plane $\plicplane$, i.e.\ $\set{\vx\in\setR^3:\iprod{\vx}{\plicnormal}-\signdist\leq0}$, equals $\refvof$ times the polyhedron volume. More formally, one may write %
$$\text{find}\quad \signdistref\in\setR\quad\text{such that}\quad\abs*{\set*{\vx\in\polyhedron*:\iprod{\vx}{\plicnormal}-\signdistref\leq0}}=\refvol.$$%
Irrespective of the spatial dimension of the polyhedron $\polyhedron*$, the interface positioning problem is one-dimensional. This property reflects the fact that the (signed) distance of two planes with a common normal $\plicnormal$, say, $\plicplane\fof{\signdist_i}$ and $\plicplane\fof{\signdist_j}$, can be computed from two arbitrary points on the respective planes, i.e.\ $\iprod{\vx-\vy}{\plicnormal}=\signdist_i-\signdist_j$ for all $(\vx,\vy)\in\plicplane\fof{\signdist_i}\times\plicplane\fof{\signdist_j}$. %
\end{note}%
%
%
\citet{JCP_1998_rvt} find the position of the PLIC plane iteratively by application of the method of \citet{brent2002}. %
%
%
An analytical relation of the volume of truncated cuboids is introduced in \citet{JCP_2000_arcl}, resorting to a piecewise definition within the respective so-called \textit{brackets}\footnote{Section~\ref{sec:mathematical_formulation} below contains a formal definition.}. A similar set of formulae was developed earlier by \citet{LANL_1993_unpublished} and implemented in \citet{AIAA_1996_vtoi}. %
%
%
\citet{JCP_2007_mmir} augmented the work of \citet{JCP_1998_rvt} by secant and bisection methods, also extending the applicability to general polyhedral meshes. However, \citet{JCP_2007_mmir} resort to a tetrahedral decomposition of the original polyhedron. %
%
%
Based on a small number of different intersection configurations, \citet{JCP_2006_arfr} and \citet{JCP_1999_vofi} derive analytical expressions for the volume of truncated tetrahedra and cuboids, respectively. %
%
%
\citet{JCP_2016_airm}, extending the work of \citet{JCP_2014_airm} from two to three spatial dimensions, propose an analytical approach to obtain the PLIC position in general convex polyhedra. After aligning the normal with the $z$-axis by applying a rotation of the coordinate system, i.e.\ $\plicnormal\mapsto\ve_z$, the polyhedron is truncated at its vertices $\vfvert_i$ by so-called\footnote{\citet{JCP_2016_airm} employ the notation $d_i$, which we have adapted to $\signdist_i$ for reasons of notational consistency.} $\signdist_i$-planes with $\signdist_i\defeq\iprod{\vfvert_i}{\plicnormal}$ and $\signdist_i\leq \signdist_{i+i}$. The intersection of $\polyhedron*$ with the negative halfspace of each of these $\signdist_i$-planes induces a volume $V_{\signdist,i}$, where $V_{\signdist,i}\leq V_{\signdist,i+1}$. Since $0\leq\refvof\leq1$, there is exactly one bracket $\mathcal{B}_i:=(\signdist_{k},\signdist_{k+1}]$ such that $\signdistref\in(\signdist_{k},\signdist_{k+1}]$ and $V_{\signdist,k}<\refvol\leq V_{\signdist,k+1}$. The corresponding planes truncate a prismatoid from the polyhedron; cf.~\reffig{diot_rotation_bracketing} for an illustration. %
A polynomial of degree three in $\signdist$, whose coefficients \citet{JCP_2016_airm} provide analytically, governs the enclosed volume, implying that the sought $\signdistref$ can be computed directly by means of root-finding. In order to obtain analytical expressions for the prismatoid volume, \citet{JCP_2016_airm} require considerable connectivity information of the underlying polyhedron a priori, along with establishing the connectivity of the intersections and cell vertices for each truncation operation. Despite the ability to considerably reduce the runtime by application of a tailored data structure, establishing connectivity increases the computational effort for complex polyhedra. 
%
%
For convex polyhedra, \citet{JCP_2018_aiir} propose an algorithm combining a tetrahedral decomposition with three root finding methods: bisection as well as the algorithms of \citet{MTOAC_1956_amfs} and \citet{brent2002}, where the latter, by design, makes no use of derivatives. By means of topological considerations of the two possible intersection configurations, the computation of the volumes of truncated tetrahedra can be carried out efficiently. A secondary, yet very important finding of \citet{JCP_2018_aiir} is the fact that the number of tetrahedra resulting from the decomposition crucially affects the overall algorithm efficiency (for the cases reported, the average number of volume computations is $6-7$). Furthermore, the fact that tetrahedral decomposition of general polyhedra poses a highly complex and demanding problem in itself, indicates the limitations of decomposition-based approaches. \citet{JCP_2018_aiir} further find that (i) significant improvement in CPU time consumption (about 50\%) can be achived by rotational transformation as used in \citet{JCP_2016_airm}. Furthermore, \citet{JCP_2018_aiir} indicate that, since the interface reconstruction usually is embedded into a larger numerical scheme, the effect of a performance gain in the interface reconstruction step becomes marginal in comparsion to the overall runtime. %
%
%
For general convex polyhedra in three spatial dimensions, \citet{JCP_2008_aagt} introduce an algorithm for volume truncation operations. In essence, they exploit the \textsc{Gaussian} divergence theorem to express the volume of the truncated polyhedron as a weighted sum of the areas of its bounding faces, the latter being  computed as a sum of outer vector products. The vertices of the truncated faces stem from both the original polyhedron and the edge intersections. For the application of the formula of \citet{GG_2004_aopp}, they require to be arranged counterclock-wise with respect to the face normal. \citet{JCP_2008_aagt} extract the aforementioned arrangement from the polyhedron topology, which quickly becomes cumbersome for polyhedra with a large number of faces. The position\footnote{\citet{JCP_2008_aagt} employ the notation $C_{\Gamma c}$, which we have adapted to $\signdistref$ for reasons of notational consistency.} $\signdistref$ is obtained by direct computation of the root of a cubic polynomial within a bracket $\mathcal{B}_i$ (cf.~section~\ref{sec:mathematical_formulation}), which requires sequentially intersecting the polyhedron with various planes parallel to $\plicplane$ passing through its vertices. Despite a carefully designed selection of the vertices, \citet{JCP_2008_aagt} state that \glqq\textit{this part of the algorithm may be the most time consuming, especially when the number of vertices of $\polyhedron*$ is large}\grqq. Since \citet{JCP_2008_aagt} do not report the average number of truncations, a meaningful comparison to the present work cannot be made. %
\citet{JCP_2016_anvc} develop a set of geometrical tools for convex polyhedra, where the PLIC positioning resorts to an algorithm of \citet{JCP_2008_aagt}. In essence, a coupling of bracketing and analytical description of the volume function
(\textbf{C}oupled \textbf{I}nterpolation-\textbf{Br}acketed \textbf{A}nalytical \textbf{V}olume \textbf{E}nforcement, CIBRAVE for short) allows to obtain the sought $\signdistref$ from spline interpolation. Each iteration $\signdist^n$, obtained from linear interpolation of $\set{\polyvof_{\mathrm{left}},\polyvof_{\mathrm{right}}}$, commences by computing the index $i$ of the parenting bracket (such that $\projvertpos{i}\leq\signdist^n\leq\projvertpos{i+1}$). Then, a truncation of the polyhedron at $\projvertpos{i}$ yields the four coefficents of the cubic spline $\mathcal{S}_i$, allowing to evaluate $\polyvof\fof{\projvertpos{i}}=\mathcal{S}_i\fof{\projvertpos{i}}$ and $\polyvof\fof{\projvertpos{i+1}}=\mathcal{S}_i\fof{\projvertpos{i+1}}$. For $\polyvof\fof{\projvertpos{i}}\leq\refvof\leq\polyvof\fof{\projvertpos{i+1}}$, the sought $\signdistref$ corresponds to the root of $\mathcal{S}_i\fof{\signdist}-\refvof$. Otherwise, the volume fractions at the bracket boundary and $\set{\polyvof_{\mathrm{left}},\polyvof_{\mathrm{right}}}$ are used to update $\signdist^n$ by linear interpolation; cf.~\refnote{cibrave_comparison} and \reffig{cibrave_step_illustration} for an illustration. %
\citet{JCP_2016_anvc} provide numerical results for tetra-, hexa-, icosa- and dodecahedra, to which we compare this work within section~\ref{sec:numerical_results}. By introducing a conceptually extensive method to extract connectivtiy from topological information, \citet{ASME_2016_aonc,JCP_2019_ncaa} extend the applicability of the previously mentioned algorithms to non-convex polyhedra, thereby nicely illustrating the complexity of establishing vertex connectivity after truncation. %
\begin{remark}[Truncation of non-convex polyhedra (\citet{JCP_2019_ncaa})]
The figure below resembles an adaption of fig.~8 from \citet{JCP_2019_ncaa}, illustrating the capability of handling non-simply connected faces. However, the proper arrangement of the polyhedron vertices and intersections, indicated by the sets on the right, requires (a) connectivity on two hierarchical levels (vertex to face/face to face) and (b) interlacing with the intersections induced by the plane $\plicplane$.\\[-1ex]%
\null\hfill%
\includegraphics{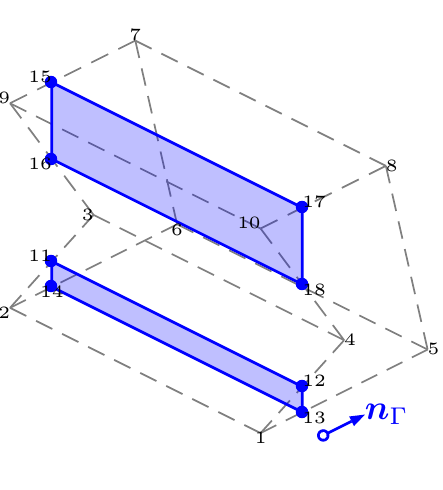}%
\hfill%
\includegraphics{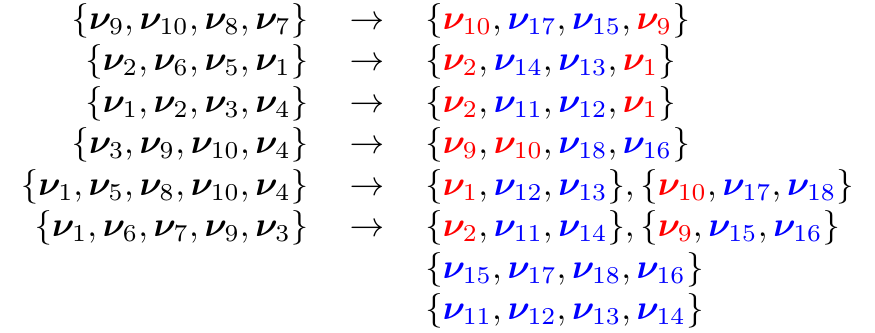}%
\hfill%
\includegraphics{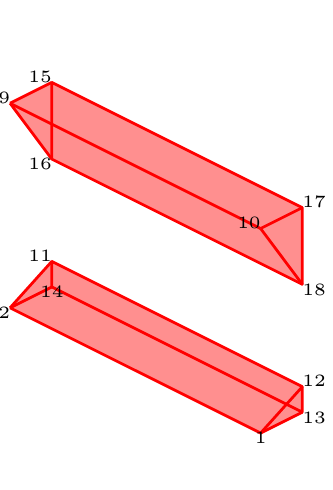}%
\hfill%
\null\\%
Even for the comparatively simple case illustrated above, figs.~9-12 in \citet{JCP_2019_ncaa} indicate the computational effort behind establishing the connectivity. %
\end{remark}%
Unfortunately, \citet{JCP_2019_ncaa} only provide a performance assessment of their algorithm in terms of relative CPU times, but not in terms of truncations, stating in their fig.~16 that the evaluation of the spline coefficients accounts for roughly 80\% of the total runtime. As a secondary finding, table~3 in \citet[p.~684]{JCP_2019_ncaa} indicates the superiority of the application of \textsc{Gaussian} divergence theorem over convex decomposition. %
%
%
Investigating an extended version of the problem under consideration here, \citet{JCP_2019_apnm} derive analytical formulae for the derivative of the parametrized volume function. The authors construct a \glqq\textit{modified Newton's method which utilizes the monotonicity of $\polyvol\fof{s}$ [notation adapted] to ensure the convergence of the iteration}\grqq. In comparison to the bisection/secant method of \citet{JCP_2007_mmir}, \citet{JCP_2019_apnm} report a reduction of iterations by 60 to 66\%. %
%
%
\citet{IJNMF_2019_aira}, following an idea of \citet{JCP_2016_airm}, apply a cell rotation to align the PLIC normal with $\ve_3$ along with a decomposition of the convex cell into $N-1$ prismatoids, where $N$ is the number of unique signed distances of the rotated vertices. The volume of the respective sub-cells is computed by exploiting the \textsc{Gaussian} divergence theorem, which allows to identify the subcell within which the sought PLIC plane must be located to obtain the prescribed volume fraction. The enclosed volume is expressed as a cubic polynomial of the normalized distance of the bounding planes, whose root is computed by a \textsc{Newton-Raphson} algorithm. The cell types under consideration in \citet{IJNMF_2019_aira} include various polyhedra with \num{12}--\num{48} vertices and \num{8}--\num{26} faces; e.g., hexagonal prisms, regular icosahedra and truncated cuboctrahedra (sic!); see~\citet[tab.~1]{IJNMF_2019_aira}. The instances of the normal $\plicnormal$ and volume fractions $\polyvof$ for the numerical experiments are arranged in one and two sets, including the potentially troublesome cases where the PLIC plane is aligned with one of the cell faces and/or the cell is almost full or empty; cf.~subsection~\ref{subsec:objective_conclusion}. The authors report very accurate reconstruction results, with both $L_2$- and $L_\infty$-type deviations of magnitude \num{e-16}. Unfortunately, \citet{IJNMF_2019_aira} do not provide average iteration numbers to which the results in section~\ref{sec:numerical_results} could be compared. %
%
%
Recently, \citet{XXX_2020_ivof} proposed a consecutive cubic spline (CCS) interpolation, which approximates the sought root $\signdistref$ by iteratively constricting $\signdist_{\mathrm{left}}<\signdistref<\signdist_{\mathrm{right}}$ such that $\brackets[s]{\signdist_{\mathrm{left}},\signdist_{\mathrm{right}}}$ maintains a sign change; cf.~\reffig{ccs_step_illustration}. Starting from $\signdist_{\mathrm{left}}\defeq\projvertposmin$ and $\signdist_{\mathrm{right}}\defeq\projvertposmax$, each iteration starts by updating %
\begin{align}
\brackets[s]{\signdist_{\mathrm{left}},\signdist_{\mathrm{right}}}\defeq%
\begin{cases}%
\brackets[s]{\signdist^n,\signdist_{\mathrm{right}}}&\vofdeviation\fof{\signdist^n}<0\\%
\brackets[s]{\signdist_{\mathrm{left}},\signdist^n}&\vofdeviation\fof{\signdist^n}>0%
\end{cases}\quad\text{with the deviation}\quad\vofdeviation\fof{\signdist}\defeq\polyvof\fof{\signdist}-\refvof.%
\end{align}%
\refFig{ccs_step_illustration} illustrates the update $\signdist^{n+1}$ as a function of the sign of $\vofdeviation\fof{\signdist^n}\vofdeviation\fof{\signdist^{n-1}}$, where, without loss of generality, it can be assumed that $\signdist^n>\signdist^{n-1}$. If the current ($\signdist^n$, $\textcolor{orange}{\blacktriangle}$) and previous ($\signdist^{n-1}$, $\textcolor{orange}{\blacksquare}$) intersection produce deviations $\vofdeviation$ with %
\begin{enumerate}%
\item[i)] \textbf{different} signs (center), they comprise the sought root $\signdistref$, i.e.\ $\signdistref\in[\signdist^{n-1},\signdist^{n}]$, such that a cubic spline interpolation based on $\vofdeviation$ and its derivative at $\signdist^n$ and $\signdist^{n-1}$, respectively, provides the next iteration ($\signdist^{n+1}$, $\textcolor{red}{\bullet}$). %
\item[ii)] \textbf{coinciding} signs (left), the sought root $\signdistref$ lies outside of $[\signdist^{n-1},\signdist^{n}]$. In this case, the next iteration ($\signdist^{n+1}$, $\textcolor{red}{\bullet}$) is calculated from the root of a linear approximation of $\vofdeviation$ around the current iteration $\signdist^n$.%
\end{enumerate}%
It is worth noting that, if $\signdist_{\mathrm{left}}$ and $\signdist_{\mathrm{right}}$ are located in the same bracket, i.e.\ $\brackets[s]{\signdist_{\mathrm{left}},\signdist_{\mathrm{right}}}\subset\brackets[s]{\signdist_{i},\signdist_{i+1}}$ (rightmost in \reffig{ccs_step_illustration}), the cubic spline \textit{exactly} interpolates the deviation $\vofdeviation$. The numerical results of \citet{XXX_2020_ivof} highlight that this is especially advantageous for $\refvof\approx0$ and $\refvof\approx1$, rendering this approach superior to \textsc{Newton}-based ones, whose convergence strongly suffers from diminishing derivatives. %

Beyond insightful comments on the involved floating point arithmetics, \citet{XXX_2020_ivof} assesses the computational performance of his algorithm in detail, concluding that, in comparison to the polyhedron truncation, the computational costs of rootfinding are negligible. Section~\ref{sec:numerical_results} contains a detailed comparison of the results. 
%
%
%
%
\subsection{Objectives}\label{subsec:objective_conclusion}%
As a consequence of the literature review, we formulate two main objectives for the present work:%
\begin{enumerate}%
\renewcommand{\labelenumi}{\textbf{O\theenumi}.}
\item Irrespective of the conceptual differences of the existing approaches, the main bottleneck for most of the published algorithms is an efficient and conceptually simple computation of the volume of truncated polyhedra. In the vast majority of the cited publications, extracting connectivity from topological information is responsible for most of the computational effort. Thus, the first objective of this work is to establish an algorithm for the computation of the volume of a truncated arbitrary polyhedron, which \textbf{restrains from extracting topological information as much as possible}. \refNote{computational_effort_topology} further discusses the computational effort of extracting toplogical information. \label{objective:topology} %
\item As has been pointed out by \citet{JCP_2018_aiir}, \citet{IJNMF_2019_aira} and \citet{XXX_2020_ivof}, the performance of a reconstruction algorithm crucially depends on the input parameters $\set{\refvof,\polyhedron*,\plicnormal}$. Hence, the numerical experiments conducted to assess the performance shall resort to a \textbf{dyadic combination of representative sets} of geometries $\polyhedron*$, volume fractions $\refvof$ and normal orientations $\plicnormal$. While \reftab{polyhedron_types} and appendix~\ref{app:supplementary_information} gather the polyhedra under consideration, the volume fractions and normals are adapted from \citet[eqs.~(25) to (28)]{XXX_2020_ivof}, i.e.
\begin{align*}
\normalset{M_{\plicnormal}}&\defeq\set*{\brackets*[s]{\cos\varphi\sin\theta,\sin\varphi\sin\theta,\cos\theta}\transpose:\brackets{\varphi,\theta}\in\frac{\pi}{2M_{\plicnormal}}\brackets[s]{1,2,\dots,2M_{\plicnormal}}\times\frac{\pi}{M_{\plicnormal}}\brackets[s]{0,1,\dots,M_{\plicnormal}}},\\%
\vofset{M_{\refvof}}&\defeq\set{10^{-k}:5\leq k\leq9}\cup\set*{10^{-4}+\frac{m-1}{M_{\refvof}-1}\brackets*{1-2\cdot10^{-4}}:1\leq m\leq M_{\refvof}}\cup\set{1-10^{-k}:5\leq k\leq9}.%
\end{align*}
We set $M_{\plicnormal}=80$ and $M_{\refvof}=50$, which corresponds to $2M_{\plicnormal}(M_{\plicnormal}+1)(M_{\refvof}+10)=\num{777600}$ numerical instances per polyhedron. Note that $\normalset{}$ corresponds to a discrete approximation of the boundary of the unit sphere, denoted $\unitsphere\defeq\set{\vx\in\setR^3:\norm{\vx}=1}$. %
Taking into account volume fractions close to zero or one becomes relevant in cases where numerical whisps occur. Here, a reconstruction tolerance of \num{e-12} is chosen. %
\end{enumerate}
%
%
\subsection{Notation \& data structure}\label{subsec:notation}%
\begin{figure}[htbp]
\null\hfill%
\subfigure[intersection with PLIC plane $\plicplane$]{\includegraphics[page=2,width=6cm]{cuboid_illustration}}%
\hfill%
\subfigure[face $\vfface_k$ with normal $\vn^\vfface_k$ and co-normals $\vN_{k,m}$]{\includegraphics[page=1,width=6cm]{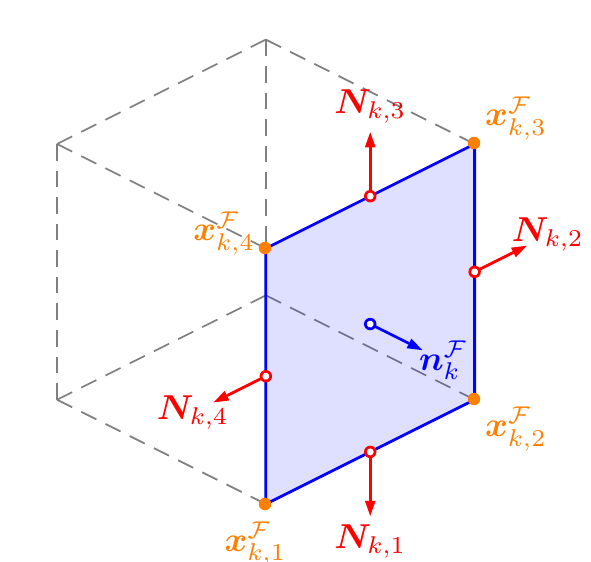}}%
\hfill%
\subfigure[interior edges $\vfedge_{k,m}^-$]{\includegraphics[page=3,width=6cm]{cuboid_illustration}}%
\hfill%
\null%
\caption{Illustration of the relevant quantities for a cuboid.}%
\label{fig:notation_illustration}%
\end{figure}

%
%
\paragraph{Geometric quantities} We consider an arbitrary polyhedron $\polyhedron*$ bounded by $N^\vfface$ planar polygonal faces $\vfface_k$ with outer unit normal $\vn_{\vfface,k}$. To ensure applicability of the \textsc{Gaussian} divergence theorem, $\partial\vfface_k$ and $\partial\polyhedron*=\bigcup_k{\vfface_k}$ are respectively assumed to admit no self-intersection\footnote{Two edges of a polygon are considered \textit{intersected} if they have a common point which is not one of the respective vertices. The analogous holds for faces.}. For the desired application within finite-volume based methods, however, self-intersecting polyhedra have no relevance. The $N^{\vfvert}_k$ vertices\footnote{Note that the number of vertices in a closed polygon coincides with the number of edges.} $\set{\vx^\vfface_{k,m}}$ on each face are ordered counter-clockwise with respect to the normal $\vn_{\vfface,k}$, implying that the $m$-th edge $\vfedge_{k,m}$ is spanned by $\brackets{\vx^\vfface_{k,m},\vx^\vfface_{k,m+1}}$. For notational convenience, assume that the indices are continued periodically, i.e.\ let $\vx^\vfface_{k,N^{\vfvert}_k+1}=\vx^\vfface_{k,1}$. To each edge $\vfedge_{k,m}$ on face $\vfface_k$ we assign the outer unit normal $\vN_{k,m}$ such that $\iprod{\vN_{k,m}}{\vn_{\vfface,k}}=\iprod{\vN_{k,m}}{\vx^\vfface_{k,m+1}-\vx^\vfface_{k,m}}=0$. \refFig{notation_illustration} illustrates the setup and its relevant quantities. %
\paragraph{Summation} Throughout this manuscript, summation over faces $\vfface_k$ and associated edges $\vfedge_{k,m}$ employ the subscripts $k$ and $m$, where the respective limits always are $k=1\dots N^\vfface$ ($m=1\dots N^{\vfvert}_k$). For ease of notation, the summation limits are omitted. %
%
%
%
\paragraph{Representation of polyhedra}%
\refTab{polyhedron_types} gathers the polyhedra employed for the numerical experiments in section~\ref{sec:numerical_results}. %
\begin{table}[htbp]%
\centering%
\caption{Polyhedra considered for the numerical investigation within this work; cf.~\refnote{parametrized_polyhedra}.}%
\label{tab:polyhedron_types}%
\begin{tabular}{c||c||c|c|c||c}%
\textbf{name}&\textbf{parameters}&\textbf{\# of faces}&\textbf{\# of edges}&\textbf{\# of vertices}&\textbf{comment}\\%
\hline%
unit cuboid(cube)&--&$6$&$12$&$8$&edge length 1\\%
(iso)dodecahedron&--&$12$&$30$&$20$&edge length 1\\%
torus&$R$,$\gamma$,$N_1$,$N_2$&$N_1N_2$&$2N_1N_2$&$N_1N_2$&cf.~appendix~\ref{app:nonstandard_polyhedron}\\%
letter A&--&$15$&$33$&$22$&cf.~appendix~\ref{app:nonstandard_polyhedron}\\%
unit tetrahedron&--&$4$&$6$&$4$&--%
\end{tabular}
\end{table}
\begin{note}[Parametrized polyhedra]\label{note:parametrized_polyhedra}%
\def\tQ{\tensor[2]{Q}}%
In principle, the classes of polyhedra given in \reftab{polyhedron_types} admit further parametrization: e.g., the normalized edge lengths of a general cuboid are $\set{1,\psi_1,\psi_2}$ with $\brackets{\psi_1,\psi_2}\subset\setR^2_+$. %
The implications of considering the parametrization becomes apparent for the tetrahedron: without loss of generality, assume that one of the vertices coincides with the origin and one of the faces (containing the aforementioned vertex) lies in the $xy$-plane. Then, any instance $\set{\polyhedron*,\refvof,\plicnormal}$ with an arbitrary (non-degenerate) tetrahedron $\polyhedron*$ can be transformed to the unit tetrahedron by a linear transformation $\tensor[2]{Q}:\setR^3\mapsto\setR^3$. While such a transformation preserves the volume fraction $\refvof$, the normal experiences a \textit{non-linear} transformation $\plicnormal\mapsto\tensor[2]{T}\fof{\plicnormal}$ with $\tensor[2]{T}\fof{\vx}\defeq\frac{\tQ\vx}{\sqrt{\iprod{\tQ\vx}{\tQ\vx}}}$. \\%
\null\hfill%
\includegraphics{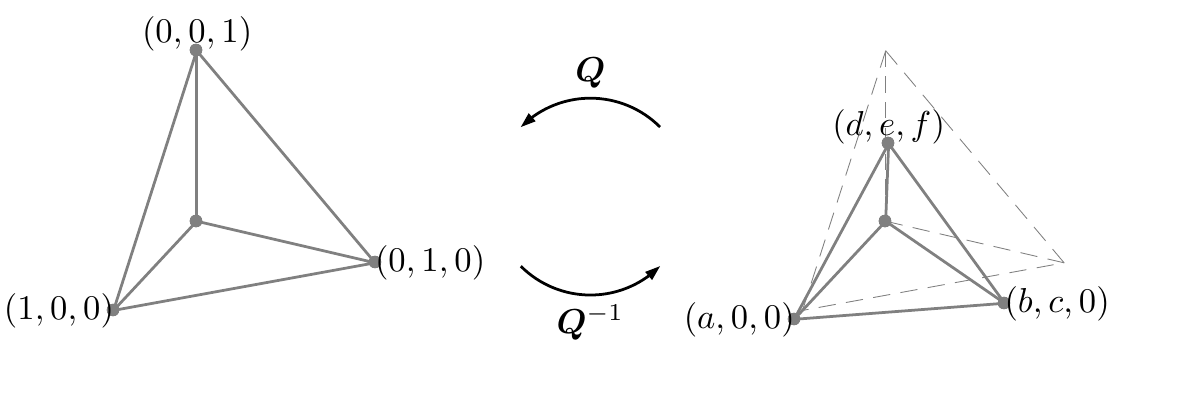}%
\hfill\null\\%
Note that the transformation $\tensor[2]{T}$ maps the boundary of the unit sphere onto itself, i.e.\ $\unitsphere=\tensor[2]{T}\fof{\unitsphere}$. Thus, if the set of normals $\normalset{}$ is dense in $\unitsphere$, it is sufficient to numerically investigate the unit tetrahedron. Since $\normalset{}$ resembles a finite approximation of the boundary of the unit sphere, this is not the case in practice; cf.~subsection~\ref{subsec:objective_conclusion}. Furthermore, even moderate deviations from the unit tetrahderon induce \textbf{strong} distortions to the sample space for the normals $\normalset{}$. Thus, the appropriate choice of $\normalset{}$ for general transformations $\tensor[2]{Q}$ poses a complex task in itself, which, however, goes beyond the scope of this work.\\%
\null\hfill%
\includegraphics[page=1]{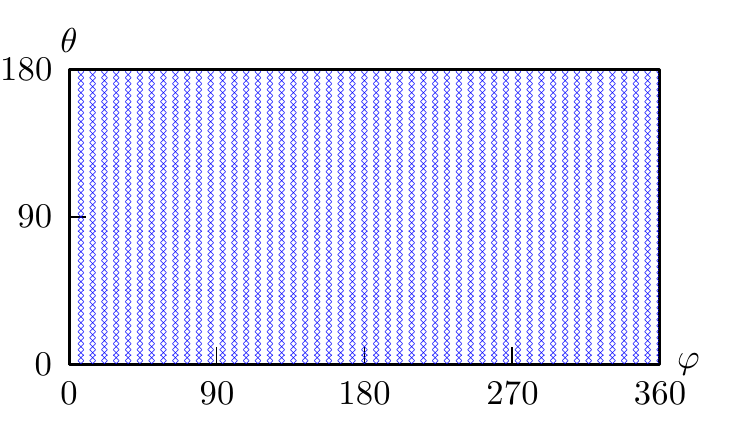}%
\hfill%
\includegraphics[page=2]{tetrahedron_normal_dist}%
\hfill\null\\%
\null\hfill%
Distortion of uniformly discretized $\unitsphere$ for $\set{\nicefrac{11}{10},\nicefrac{1}{2},\nicefrac{9}{10},-\nicefrac{1}{4},-\nicefrac{1}{10},\nicefrac{3}{10}}$, corresponding to the figure above.%
\hfill\null%
\end{note}

%
%
%
In algorithmic terms, the following two data sets represent a polyhedron: 
\begin{enumerate}
\item a list of $N^{\vfvert}$ \textbf{common vertices} $\set{\vfvert_i}$. (\textit{In theory, every face $\vfface_k$ could hold its own set of vertices.}) 
\item a list of $N^\vfface$ faces $\set{\vfface_k}$, each defined by a \textbf{periodic set of indices} $\set{i_m}$ with $1\leq m\leq N^{\vfvert}_k$. The corresponding vertices form a closed \textbf{planar} polygon, ordered counter-clockwise with respect to the face normal $\vn_{\vfface,k}$. (\textit{Here, by design, a hole in a face is realized by an additional coplanar face with inverted orientation; cf.~\reffig{polygon_hole_definition}.})%
\begin{figure}[htbp]%
\null\hfill%
\includegraphics[page=1]{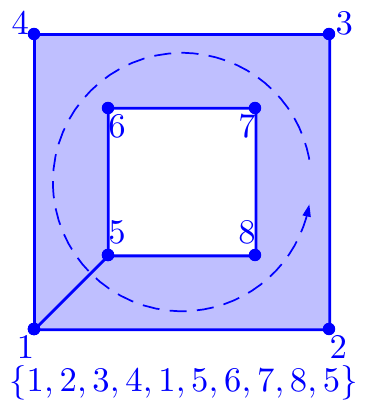}%
\hfill%
\includegraphics[page=2]{polygon_hole_definition}%
\hfill\null%
\caption{Different strategies to define a closed polygon with a hole: single polygon with overlapping edges ($\{1,5\}$ and $\{5,1\}$) and decomposition with appropriate ordering.}%
\label{fig:polygon_hole_definition}%
\end{figure}
\end{enumerate}
%
%
\begin{note}[Computational effort vs.\ topology]\label{note:computational_effort_topology}%
The above data sets comprise the minimum information required to compute the volume of a polyhedron. In fact, most unstructured meshes employed in VOF-codes resort to this type of description. In theory, one could deduce a list of edges, each respectively shared by two faces. However, this connectivity has to be either (i) provided a priori (increasing memory consumption) or (ii) established at runtime (increasing CPU consumption), where dynamically changing meshes cannot resort to (i). At this point, the authors would like to emphasize that, obviously, the computational effort of volume computation could be reduced roughly by half, since neglecting the edge connectivity implies that the number of edge operations (intersections) doubles with the number of faces $N^\vfface$. On the other hand, the complexity of establishing the edge-connectivity between faces increases \textbf{quadratically} with $N^\vfface$, i.e., it becomes relatively expensive, even for a moderate number of faces $N^\vfface$. Thus, we pursue a \textbf{face-based} design, exclusively resorting to intra-facial connectivity. 
\end{note}%
%
%
%
%
\section{Mathematical formulation}\label{sec:mathematical_formulation}%
\begin{figure}[htbp]
\centering%
\includegraphics{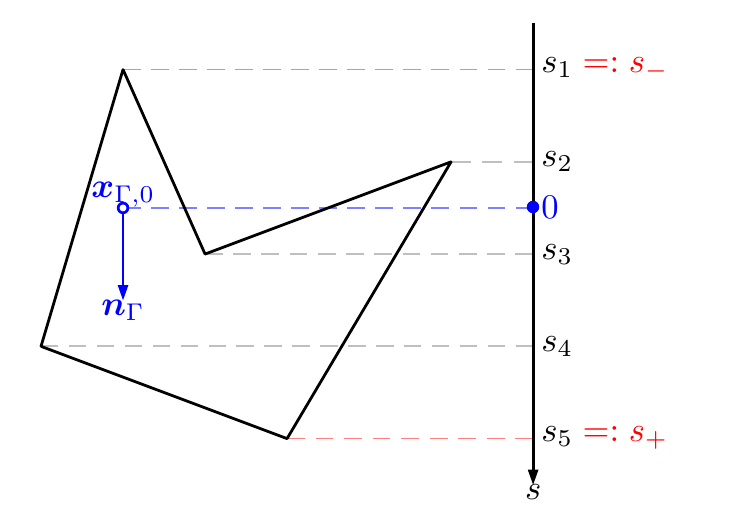}%
\caption{Polygonal face with signed distances of the vertices.}%
\label{fig:vertex_normal_projections}%
\end{figure}

For a given polyhedron $\polyhedron*$ and a unit normal $\plicnormal$, we wish to compute a base point $\plicbaseref$ for a plane $\plicplane\fof{\plicbase}$ with normal $\plicnormal$, such that the intersection of the negative halfspace of $\plicplane$ with $\polyhedron*$, denoted %
\begin{align}
\polyhedron\fof{\plicbase}\defeq\set{\vx\in\polyhedron*:\iprod{\vx-\plicbase}{\plicnormal}\le0},\label{eqn:polyhedron_intersection}%
\end{align}
admits some given positive volume $\refvol\leq\abs{\polyhedron*}$, where henceforth $\polyvol\fof{\plicbase}\defeq\abs{\polyhedron\fof{\plicbase}}$ denotes the volume of the truncated polyhedron. Note that $\refvol=\polyvol\fof{\plicbaseref}$. %
With an arbitrary (but spatially fixed) reference $\xbase$, the \textit{signed distance} $\signdist$ is used to parametrize the base point in \refeqn{polyhedron_intersection} as $\plicbase=\xbase+s\plicnormal$, implying %
\begin{align}
\plicplane\fof{\plicbase}=\plicplane\fof{\signdist;\xbase}=\set{\vx\in\setR^3:\pliclvlset\fof{\vx;\signdist,\xbase}=0}\quad\text{with}\quad\pliclvlset\fof{\vx;\signdist,\xbase}\defeq\iprod{\vx-\xbase}{\plicnormal}-\signdist.\label{eqn:plic_parametrization_levelset}%
\end{align}
Then, we obtain the problem formulated in \refnote{problem_formulation}, i.e. %
\begin{align}
\text{find}\quad\signdistref\in\setR\quad\text{such that}\quad\abs*{\set*{\vx\in\polyhedron*:\iprod{\vx-\xbase}{\plicnormal}-\signdistref\leq0}}%
=\refvol.\label{eqn:positioning_problem_formulation}%
\end{align}
In what follows, a prime denotes the derivative with respect to the signed distance $\signdist$. %
\paragraph{Non-dimensional formulation.} Throughout this work, we shall resort to the normalized volume or \textit{volume fraction} $\polyvof\fof{\signdist}\defeq\polyvol\fof{\signdist}\abs{\polyhedron*}^{-1}$. The signed distances of the vertices
\begin{align}
\hat{\mathcal{S}}
\defeq\bigcup_k\set*{\iprod{\vx^\vfface_{k,m}-\xbase}{\plicnormal}}\label{eqn:vertex_normal_projections}%
\end{align}
correspond to the values of $\signdist$ for which the plane $\plicplane$ passes through a vertex (single or multiple) of the polyhedron $\polyhedron*$. The elements $\projvertpos{i}$ of $\hat{\mathcal{S}}$, being unique ($s_i\neq s_j$ for $i\neq j$) and arranged in ascending order ($\projvertpos{i}<\projvertpos{i+1}$), define mutually disjoint so-called \textit{brackets} $\mathcal{B}_i\defeq(\projvertpos{i},\projvertpos{i+1}]$ with $\mathcal{B}_1\defeq[s_1,s_2]$. The \textit{target} bracket $\refbracket$ contains the sought root $\signdistref$; cf.~\reffig{vertex_normal_projections} for an illustration in two spatial dimensions. %
Furthermore, let %
\begin{align}
\tilde{\signdist}\defeq\frac{\signdist-\projvertposmin}{\projvertposmax-\projvertposmin}%
\quad\text{with}\quad\projvertposmin\defeq\min_i\projvertpos{i},\quad\projvertposmax\defeq\max_i\projvertpos{i}%
\quad\text{and}\quad\projvertsize\defeq\projvertposmax-\projvertposmin,%
\end{align}
where, henceforth, the tilde denotes the normalized pendant of a variable. Note that $\partial_{\tilde{\signdist}}=\projvertsize\partial_s$. %
\paragraph{Regularity.} Globally, $\polyvol\fof{\signdist}\in\mathcal{C}^0$ is a strictly increasing function, whose derivative corresponds to the area of the PLIC plane contained in the computational cell\footnote{This identity directly results from application of the \textsc{Reynolds} transport theorem to \refeqn{polyhedron_intersection_volume}. However, \citet[eq.~(30)]{JCP_2019_apnm} offer an intuitive geometric derivation.}, i.e.\ $\polyvol^\prime\fof{\signdist}=\abs{\polyhedron*\cap\plicplane\fof{\signdist}}\geq0$. Locally, i.e.\ within a bracket $\mathcal{B}_i$, the volume is a third-order polynomial in $\signdist$. %
Under mild restrictions concerning the PLIC normal $\plicnormal$, cf.~\refnote{boundary_value_volume_derivative}, the truncated volume is globally continuously differentiable, i.e.\ $\polyvol\fof{\signdist}\in\mathcal{C}^1$. 
\begin{note}[Faces parallel to $\plicplane$]\label{note:boundary_value_volume_derivative}%
Consider a face $\vfface_k$ which is parallel to the PLIC plane $\plicplane$, i.e.\ $\pm1=\iprod{\vn^\vfface_k}{\plicnormal}\defeq*\sigma_k$. Hence, depending on the value of $\signdist$, the face is either entirely contained in the PLIC plane or not at all, corresponding to a multivalued (in the sense of non-smooth analysis; see, e.g., \citet{nonsmooth_analysis}) derivative $\polyvof^\prime\fof{\signdist}$ for the respective parameters. %
\begin{center}
\null\hfill%
\includegraphics[page=1]{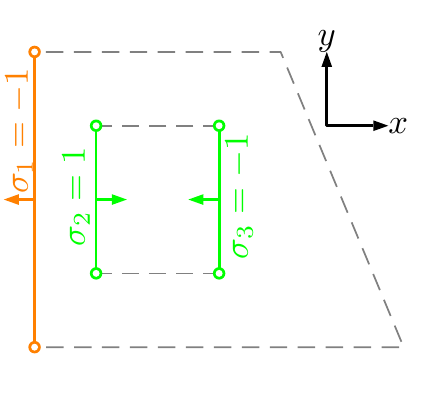}%
\hfill%
\includegraphics[page=2]{area_boundary_jump_illustration}%
\hfill%
\includegraphics[page=4]{area_boundary_jump_illustration}%
\hfill%
\includegraphics[page=1]{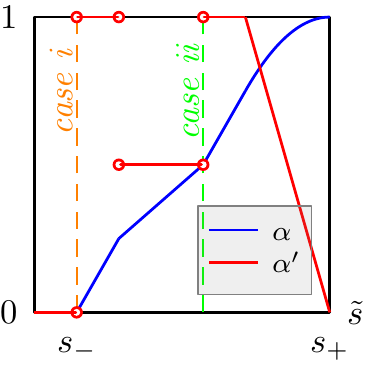}%
\hfill\null%
\end{center}
Let $\mathcal{H}_{\polyhedron*}$ denote the \textbf{convex hull} of the polyhedron and $\partial\mathcal{H}_{\polyhedron*}$ its boundary, with $\partial\mathcal{H}_{\polyhedron*}=\partial\polyhedron*=\bigcup_k{\vfface_k}$ for convex polyhedra. The regularity of $\polyvof\fof{\signdist}$ depends on the topological properties of the respective face $\vfface_k$: %
\begin{enumerate}%
%
%
\item[i.] If $\vfface_k$ is an element of $\partial\mathcal{H}_{\polyhedron*}$ ({\color{orange}\rule[2pt]{10pt}{1pt}} in the leftmost figure), the corresponding signed distance is $\projvertposmin$ ($\sigma_k=-1$) or $\projvertposmax$ ($\sigma_k=1$). These boundary-discontinuities of $\polyvof^\prime\fof{\signdist}$ can be resolved by assigning the respectively interior limit (right-sided for $\sigma_k=-1$ and left-sided for $\sigma_k=1$). Thus, let %
\begin{align}
{\color{orange}\polyvof^\prime\fof{\projvertposmin}\defeq\lim\limits_{s\searrow\projvertposmin}\polyvof^\prime\fof{\signdist}=\abs{\polyhedron*}^{-1}\sum\limits_k{\abs{\vfface_k\cap\plicplane\fof{\projvertposmin}}}}%
\quad\text{and}\quad%
\polyvof^\prime\fof{\projvertposmax}\defeq\lim\limits_{s\nearrow\projvertposmax}\polyvof^\prime\fof{\signdist}=\abs{\polyhedron*}^{-1}\sum\limits_k{\abs{\vfface_k\cap\plicplane\fof{\projvertposmax}}},%
\end{align}
where the sum accounts for multiple coplanar faces. %
%
%
\item[ii.] A face $\vfface_k$ which is \textbf{not} an element of $\partial\mathcal{H}_{\polyhedron*}$ ({\color{green}\rule[2pt]{10pt}{1pt}} in the leftmost figure) induces an \textbf{interior} jump discontinuity (\textcolor{red}{$\circ$} in the rightmost figure) in $\polyvof^\prime\fof{\signdist}$, where $\polyvof^{\prime\prime}\fof{\signdist}$ may jump as well if $\plicplane$ simultaneously passes through a vertex not contained in $\vfface_k$. The third figure above illustrates the multi-valued derivative as $\polyvof^\prime\in[\nicefrac{1}{2},1]$ at $\polyvof=\nicefrac{1}{2}$. %
\end{enumerate}
Moreover, the absence of faces parallel to $\plicplane$ implies vanishing volume derivatives at the boundaries, i.e.
\begin{align}
\set{\sigma_k}\not\ni\pm1\quad\implies\quad\polyvof^\prime\fof{\projvertposminmax}=0,\label{eqn:volume_derivative_boundary_value}%
\end{align}
which can be assumed to be the case in the vast majority of applications. 
Hence, any convex polyhedron always admits a globally continuously differentiable volume function, i.e.\ it holds that $\polyvof\fof{\signdist}\in\mathcal{C}^1$ for all $\projvertposmin<\signdist<\projvertposmax$, irrespective of the alignment of $\plicnormal$ with the normals of the faces. For non-convex polyhedra, the regularity depends on $\set{\sigma_k}$. %
\end{note}
The desired numerical treatment of \refeqn{positioning_problem_formulation} suggests a non-dimensional formulation. Thus, we define the volume fraction deviation %
\begin{align}
\vofdeviation\fof{\signdist}\defeq\polyvof\fof{\signdist}-\refvof%
\quad\text{with}\quad%
\vofdeviation:[\projvertposmin,\projvertposmax]\mapsto[-\refvof,1-\refvof].%
\end{align}
Since the volume fraction $\polyvof$ is (at least) continuous and strictly increasing, $\signdistref$ corresponds to the unique root of the deviation $\vofdeviation$ in $[\projvertposmin,\projvertposmax]$, whose approximation we compute iteratively by %
\begin{align}
\signdist^{n+1}=\signdist^n+\Delta\fof{\signdist^n}.%
\label{eqn:position_minimum_newton_iteration}%
\end{align}
The computation of the initial value $\signdist^0$ and step size $\Delta\signdist^n$, respectively, are deferred to subsections~\ref{subsec:initial_value} and \ref{subsec:stepsize}. The upcoming section derives the expressions for the truncated volume $\polyvol\fof{\signdist}$ and its derivatives. %
%
%
\subsection{Volume computation}\label{subsec:volume_computation}%
\begin{figure}[htbp]%
\null\hfill%
\includegraphics[page=1]{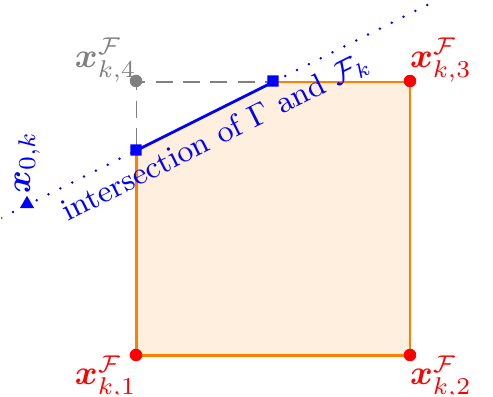}
\hfill%
\includegraphics[page=2]{face_area_example}
\hfill%
\includegraphics[page=3]{face_area_example}
\hfill\null%
\caption{Illustration of the coefficients for the face area computation given in \refeqnsave{face_coefficients}. Note that, due to the choice of $\vx_{0,k}$ in \refeqnsave{face_intersection_base_point}, the boundary segment located on $\plicplane$ (right) does not contribute to the area integral.}%
\label{fig:face_area_illustration}%
\end{figure}

By application of the \textsc{Gaussian} divergence theorem, the volume of the truncated polyhedron $\polyhedron$ can be written as
\begin{align}
\polyvol=\intpoly{1}
=\frac{1}{3}\intdpoly{\iprod{\vx-\xref}{\vn_{\partial\polyhedron}}}=%
\frac{1}{3}\brackets*[s]{\intdpolynoplic{\iprod{\vx-\xref}{\vn_{\partial\polyhedron}}}+\intplic{\iprod{\vx-\xref}{\plicnormal}}}\label{eqn:polyhedron_intersection_volume},%
\end{align}
where the choice of the reference $\xref$ is arbitrary\footnote{In the literature, the common choice is $\xref\equiv\vec{0}$; see, e.g., \citet{JCP_2008_aagt,JCP_2016_anvc,JCP_2019_ncaa}.}. %
The fact that $\plicbase\in\plicplane$ suggests to choose $\xref\defeq\xbase+\signdist\plicnormal$, in which case the integral over $\plicplane$ vanishes and \refeqn{polyhedron_intersection_volume} simplifies to
\begin{align}
\polyvol\fof{\signdist}=\frac{1}{3}\intdpolynoplic{\iprod{\vx-\xbase-\signdist\plicnormal}{\vn_{\partial\polyhedron}}}.\label{eqn:polyhedron_intersection_volume_simplified}%
\end{align}
\begin{remark}[Invariance under translation/rotation]%
\def\tQ{\tensor[2]{Q}}%
Any affine linear transformation $\tT:\vx\mapsto\tQ\transpose\vx+\xref$ with $\det{\tQ}=1$ and $\tQ\transpose\tQ=\ident$ preserves the volume of a bounded domain $\domain\subset\setR^3$, implying that the volume of $\domain$ is invariant under translation and rotation. The integrand in the rightmost expression of \refeqn{polyhedron_intersection_volume} corresponds to the local signed distance (which is invariant under rotation) of the boundary to the translated origin $\xref$. Since the boundary $\partial\polyhedron$ is composed of planar segments, the signed distance is constant within each segment and, hence, can be extracted from the integral, leaving to compute the area of the respective segment $\vfface_k^-$. If the signed distance to the origin is zero, the respective segment does not contribute to the volume integral. Thus, if the origin is translated to be coplanar to $\plicplane$ (note that $\xref\in\polyhedron*\cap\plicplane$ is not required), one obtains $\int_{\plicplane}{\iprod{\vx-\xref}{\plicnormal}\darea}\equiv0$. %
\end{remark}
Note that the integration domain in the above expression is piecewise planar, with, say, $\partial\polyhedron\setminus\plicplane\defeq*\bigcup_k\vfface_k^-$, where $\vfface_k^-=\vfface_k^-\fof{\signdist}$ denotes the segment of $\vfface_k$ located in the negative halfspace of $\plicplane\fof{\signdist}$; cf.~\reffig{notation_illustration}. %
\refEqn{polyhedron_intersection_volume_simplified} becomes %
\begin{align}
\polyvol\fof{\signdist}%
=\frac{1}{3}\sum\limits_{k}{\iprod{\vx^\vfface_{k,1}-\xbase-\signdist\plicnormal}{\vn_{\vfface,k}}\int\limits_{\vfface_k^-}{1\darea}}%
=\frac{1}{3}\sum\limits_{k}{\brackets{\volcoeffconst{k}+\signdist\volcoefflin{k}}\int\limits_{\vfface_k^-}{1\darea}}=\frac{1}{3}\sum\limits_{k}{\brackets{\volcoeffconst{k}+\signdist\volcoefflin{k}}A_k\fof{\signdist}},\label{eqn:polyhedron_volume_coeff_face}%
\end{align}
where the coefficients
\begin{align}
\volcoeffconst{k}\defeq\iprod{\vx^\vfface_{k,1}-\xbase}{\vn_{\vfface,k}}\qquad\text{and}\qquad\volcoefflin{k}\defeq-\iprod{\vn_{\vfface,k}}{\plicnormal}\label{eqn:volume_coefficients}%
\end{align}
depend only on static geometric quantities\footnote{\textit{Static} in the sense of not depending on the position of $\plicplane$, i.e., on $\signdist$.}, which is advantageous in terms of computational efficiency. Note that \refeqn{polyhedron_volume_coeff_face} corresponds to \refeqn{polyhedron_intersection_volume}, where the spatial dimension is reduced by one. Applying the \textsc{Gauss} theorem once more, the area $A_k\defeq\abs{\vfface_k^-}$ of a planar face can be written as %
\begin{align}
A_k\fof{\signdist}=\int\limits_{\vfface_k^-}{1\darea}=\frac{1}{2}\brackets*[s]{\int\limits_{\partial\vfface_k^-\setminus\plicplane}{\iprod{\vx-\vx_{0,k}\fof{\signdist}}{\vN_{\partial\vfface_k}}\dline}+\int\limits_{\partial\vfface_k^-\cap\plicplane}{\iprod{\vx-\vx_{0,k}\fof{\signdist}}{\vN_{\partial\vfface_k}}\dline}}.\label{eqn:intersected_face_area}%
\end{align}
\begin{remark}[Area computation]%
In one form or another, most literature contributions, e.g., \citet{JCP_2008_aagt}, apply the formula of \citet{GG_2004_aopp} to compute the area of a planar polygon:\\[10pt]%
\null\hspace{.5cm}%
\includegraphics[page=1]{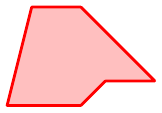}%
\multido{\i=2+1}{4}{\hspace{.25cm}\ifthenelse{\equal{\i}{2}}{\raisebox{.5cm}{=}}{\raisebox{.5cm}{+}}\hspace{.25cm}%
\includegraphics[page=\i]{goldman_cross_product}%
}%
\null\\%
Thereto, its vertices -- potentially containing intersections of $\partial\vfface_k$ with $\plicplane$ -- must be ordered counter-clockwise with respect to $\vn_{\vfface,k}$, imposing additional computational effort. On the other hand, the application of the \textsc{Gaussian} divergence theorem requires no connectivity beyond the a priori assignment of vertices to edges: \\[10pt]%
\null\hspace{.5cm}%
\includegraphics[page=1]{goldman_cross_product}%
\multido{\i=6+1}{6}{\hspace{.25cm}\ifthenelse{\equal{\i}{6}}{\raisebox{.5cm}{=}}{\raisebox{.5cm}{+}}\hspace{.25cm}%
\includegraphics[page=\i]{goldman_cross_product}%
}%
\null%
\end{remark}
In order to eliminate the last summand in \refeqn{intersected_face_area} as above, one needs to find a point $\vx_{0,k}\in\plicplane$ with zero normal distance to $\vfface_k$, i.e.\ such that $\iprod{\vx_{0,k}-\vx^\vfface_{k,1}}{\vn^\vfface_k}=0$. Note that the latter does not imply that $\vx_{0,k}$ is an element of $\vfface_k^-$ (nor, in fact, of $\vfface_k$). %
If the PLIC plane $\plicplane$ and the face $\vfface_k$ are not parallel, i.e.\ for $\abs{\iprod{\vn^\vfface_k}{\plicnormal}}<1$, their line of intersection (cf.~\reffig{face_area_illustration} for an illustration) contains
\begin{align}
\vx_{0,k}\fof{\signdist}=%
\frac{\iprod{\xbase}{\plicnormal}+s-\iprod{\vx^\vfface_{k,1}}{\vn^\vfface_k}\iprod{\vn^\vfface_k}{\plicnormal}}{1-\iprod{\vn^\vfface_k}{\plicnormal}^2}\plicnormal+%
\frac{\iprod{\vx^\vfface_{k,1}}{\vn^\vfface_k}-\brackets{\iprod{\xbase}{\plicnormal}+s}\iprod{\vn^\vfface_k}{\plicnormal}}{1-\iprod{\vn^\vfface_k}{\plicnormal}^2}\vn^\vfface_k.\label{eqn:face_intersection_base_point}%
\end{align}
As a beneficial side-effect, this choice of the reference point, compared to, say, the origin, reduces the numerical errors; cf.~\citet{GeometricRobustness2013}. %
Decomposing the boundary as $\partial\vfface_k^-\setminus\plicplane\defeq*\bigcup_m\vfedge_{k,m}^-$ and inserting the base point from \refeqn{face_intersection_base_point}, \refeqn{intersected_face_area} becomes
\begin{align}
A_k\fof{\signdist}=\frac{1}{2}\sum\limits_{m}{\iprod{\vx^\vfface_{k,m}-\vx_{0,k}\fof{\signdist}}{\vN_{k,m}}\int\limits_{\vfedge_{k,m}^-}{1\dline}}.%
\end{align}
Introducing $l_{k,m}\fof{\signdist}\defeq\int_{\vfedge_{k,m}^-}{1\dline}$ for notational convenience, one obtains %
\begin{align}
A_k\fof{\signdist}=\frac{1}{2}\sum\limits_{m}{\brackets*{\areacoeffconst{k}{m}+\signdist\areacoefflin{k}{m}}l_{k,m}\fof{\signdist}}\label{eqn:intersected_face_area_reduced}%
\end{align}
with the coefficients
\begin{align}
\areacoeffconst{k}{m}\defeq\iprod{\vx^\vfface_{k,m}}{\vN_{k,m}}+\frac{\iprod{\vx^\vfface_{k,1}}{\vn^\vfface_k}\iprod{\vn^\vfface_k}{\plicnormal}-\iprod{\xbase}{\plicnormal}}{1-\iprod{\vn^\vfface_k}{\plicnormal}^2}\iprod{\vN_{k,m}}{\plicnormal}%
\quad\text{and}\quad%
\areacoefflin{k}{m}\defeq-\frac{\iprod{\vN_{k,m}}{\plicnormal}}{1-\iprod{\vn^\vfface_k}{\plicnormal}^2}.\label{eqn:face_coefficients}%
\end{align}
As in \refeqn{volume_coefficients}, the coefficients depend only on static geometric quantities, implying that their evaluation is required only once per cell. %
Note that $l_{k,m}$, corresponding to the length of the segment of the edge $\vfedge_{k,m}$ located within the negative halfspace of $\plicplane$, is a positive piecewise linear function of $\signdist$, implying that the derivative $l_{k,m}^\prime$ is a positive constant, once again depending only on static geometric quantities. %

Finally, combining \refeqn**{intersected_face_area_reduced} and \refeqn*{polyhedron_volume_coeff_face} yields the truncated volume
\begin{align}
\polyvol\fof{\signdist}=\frac{1}{6}\sum\limits_{k}{\brackets{\volcoeffconst{k}+\signdist\volcoefflin{k}}\sum\limits_{m}{\brackets*{\areacoeffconst{k}{m}+\signdist\areacoefflin{k}{m}}l_{k,m}\fof{\signdist}}},\label{eqn:polyhedron_volume_signed_distance}%
\end{align}
which is a continuous function composed of third-order polynomials in $\signdist$; cf.~\refnote{boundary_value_volume_derivative}. %
Conceptually, the above equation can be interpreted as follows: by recursively applying the \textsc{Gaussian} divergence theorem, the volume integral reduces to a set of line integrals (corresponding to the computation of the interior edge lengths $l_{k,m}$), while the static coefficients $\set{\volcoeffconst{k},\volcoefflin{k},\areacoeffconst{k}{m},\areacoefflin{k}{m}}$ contain the required topological information. %
The derivatives of \refeqn{polyhedron_volume_signed_distance} are %
\begin{align}
\begin{split}%
\polyvol^\prime\fof{\signdist}&=\frac{1}{3}\sum\limits_{k}{\brackets{\volcoeffconst{k}+\signdist\volcoefflin{k}}A_k^\prime\fof{\signdist}+\volcoefflin{k}A_k\fof{\signdist}},\\%
\polyvol^{\prime\prime}\fof{\signdist}&=\frac{1}{3}\sum\limits_{k}{\brackets{\volcoeffconst{k}+\signdist\volcoefflin{k}}A_k^{\prime\prime}+2\volcoefflin{k}A_k^\prime\fof{\signdist}}\quad\text{and}\quad
\polyvol^{\prime\prime\prime}\fof{\signdist}=\sum\limits_{k}{\volcoefflin{k}A_k^{\prime\prime}}%
\end{split}%
\end{align}%
with the face area derivatives %
\begin{align}
A_k^\prime\fof{\signdist}=\frac{1}{2}\sum\limits_{m}{\brackets*{\areacoeffconst{k}{m}+\signdist\areacoefflin{k}{m}}l_{k,m}^\prime+\areacoefflin{k}{m}l_{k,m}\fof{\signdist}}%
\quad\text{and}\quad%
A_k^{\prime\prime}=\sum\limits_{m}{\areacoefflin{k}{m}l_{k,m}^\prime}.\label{eqn:intersected_face_area_reduced_derivative}%
\end{align}
%
%
\begin{remark}[Comparison to \citet{JCP_2016_anvc}]
In their eq.~(17), \citet{JCP_2016_anvc} obtain a representation of the volume similar to \refeqn{polyhedron_volume_signed_distance}, say $$\polyvol_{i}\fof{\signdist}=\beta_{3,i}\signdist^3+\beta_{2,i}\signdist^2+\beta_{1,i}\signdist+\beta_{0,i}\quad\text{for}\quad\signdist\in[\signdist_i,\signdist_{i+1}]$$%
by truncating the polyhedron at either $\projvertpos{i}$ or $\projvertpos{i+1}$. However, the complex geometric operations involved imply a considerable increment in computational effort. Noting that the highest non-trivial derivative of \refeqn{polyhedron_volume_signed_distance} is $\polyvol^{\prime\prime\prime}\fof{\signdist}$, being piecewise constant within a bracket $[\signdist_i,\signdist_{i+1}]$, yields the polynomial coefficients as a by-product of a single truncation at an \textbf{arbitrary} $\signdist\in[\signdist_i,\signdist_{i+1}]$; cf.~subsection~\ref{subsubsec:implicit_bracketing}. %
\end{remark}%

The robust numerical computation of the lengths $l_{k,m}$ and, consecutively, the areas $A_k$, along with their derivatives, requires considering the topological properties of the involved entities, which shall be addressed in the upcoming subsection. %
%
%
\subsection{Topological properties of geometric entities}\label{subsec:topological_considerations}%
The decomposition of a non-degenerately (\refnote{edgeface_on_plic} considers the degenerate case) intersected face $\vfface_k$ requires a hierarchically consistent evaluation of the topological properties of its vertices $\set{\vx^\vfface_k}$ and edges $\set{\vfedge_{k,m}}$. With respect to the orientable hypersurface $\plicplane$, described by the level-set function given in \refeqn{plic_parametrization_levelset}, the logical status $\vfstatus{}$ of any of a vertex is either interior ($\vfstatus{}=-1$), intersected ($\vfstatus{}=0$) or exterior ($\vfstatus{}=1$). %
\begin{impldetail}[Robustness]%
For the purpose of numerical robustness, the status assignment employs a tubular neighborhood of thickness $2\zerotol$ around $\plicplane$, corresponding to the interval $(-\zerotol,\zerotol)$, i.e.%
\begin{align}%
\vfstatus{\vx^\vfface_k}\defeq%
\begin{cases}%
\phantom{-}0&\text{if}\quad\abs{\pliclvlset\fof{\vx^\vfface_k}}<\zerotolerance\\%
\sign{\pliclvlset\fof{\vx^\vfface_k}}&\text{if}\quad\abs{\pliclvlset\fof{\vx^\vfface_k}}\geq\zerotolerance%
\end{cases}.%
\end{align}
In other words, any point whose absolute distance to $\plicplane$ falls below $\zerotol$ is considered to be on $\plicplane$. The choice of an appropriate tolerance strongly depends on the absolute value of the characterisic length scale $h$ of the polyhedron, especially for $h\ll1$. Throughout this work, we have $h\approx1$ and, thus, let $\zerotolerance\defeq\num{e-14}$. In fact, all zero-comparisons are implemented in this way. %
\end{impldetail}%
The hierarchically superior entity is an edge $\vfedge$, whose status is a surjective function of the status of its associated vertices $\set{\vfvert_i,\vfvert_j}$. %
\begin{table}[htbp]
\centering%
\caption{Edge status as function of the stat\={u}s of the associated vertices; cf.~\refnote{edgeface_on_plic}.}%
\label{tab:intersection_status_edge}%
\begin{tabular}{c||ccc|cc|cc|c}
&\multicolumn{3}{c|}{}&\multicolumn{5}{c}{\textbf{degenerate}}\\%
&exterior&interior&intersected&\multicolumn{2}{c|}{exterior}&\multicolumn{2}{c|}{interior}&intersected\\%
&\includegraphics[page=4]{edgestatus_illustration}&\includegraphics[page=2]{edgestatus_illustration}&\includegraphics[page=1]{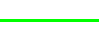}&\multicolumn{2}{c|}{\includegraphics[page=5]{edgestatus_illustration}}&\multicolumn{2}{c|}{\includegraphics[page=3]{edgestatus_illustration}}&\includegraphics[page=6]{edgestatus_illustration}\\%
\hline%
$\vfstatus{\vfvert_i}$&$1$&$-1$&$\pm1$&$1$&$0$&$-1$&$\phantom{-}0$&$0$\\%
$\vfstatus{\vfvert_j}$&$1$&$-1$&$\mp1$&$0$&$1$&$\phantom{-}0$&$-1$&$0$\\%
\hline%
$\vfstatus{\vfedge}$&$1$&$-1$&$0$&\multicolumn{2}{c|}{$2$}&\multicolumn{2}{c|}{$-2$}&$3$%
\end{tabular}%
\begin{tabular}{c}
\includegraphics[page=7]{edgestatus_illustration}%
\end{tabular}
\end{table}%
An edge with at least one of its vertices located on $\plicplane$ becomes degenerate, where the type of degeneration depends on the status of the respectively complementary vertex; cf.~\reftab{intersection_status_edge}. 
%
%
The length of an interior/exterior edge $\vfedge_{k,m}$ (including the respective degenerate cases) with vertices $\set{\vfvert_i,\vfvert_j}$ is trivially computed as
\begin{align}
l_{k,m}=%
\begin{cases}%
0&\text{if}\quad\vfstatus{\vfedge_{k,m}}\in\set{1,2},\\%
\norm{\vfvert_i-\vfvert_j}&\text{if}\quad\vfstatus{\vfedge_{k,m}}\in\set{-1,-2}%
\end{cases}.\label{eqn:edgelength}%
\end{align}
For an intersected edge ($\vfstatus{\vfedge_{k,m}}=0$), one obtains the monotonous piecewise linear function %
\begin{align}
l_{k,m}\fof{\signdist}=%
\begin{cases}
\frac{\iprod{\vfvert_i-\xbase}{\plicnormal}-s}{\iprod{\vfvert_i-\vfvert_j}{\plicnormal}}\norm{\vfvert_i-\vfvert_j}&\text{if}\quad\vfstatus{\vfvert_i}=-1,\\%
\frac{\iprod{\vfvert_j-\xbase}{\plicnormal}-s}{\iprod{\vfvert_j-\vfvert_i}{\plicnormal}}\norm{\vfvert_i-\vfvert_j}&\text{if}\quad\vfstatus{\vfvert_j}=-1,%
\end{cases}%
\end{align} 
whose derivative with respect to the signed distance $\signdist$, i.e. %
\begin{align}
l^\prime_{k,m}=%
\begin{cases}
0&\text{if}\quad\vfstatus{\vfedge_{k,m}}\in\set{\pm1,2},\\%
\frac{\norm{\vfvert_i-\vfvert_j}}{\abs{\iprod{\vfvert_i-\vfvert_j}{\plicnormal}}}&\text{if}\quad\vfstatus{\vfedge_{k,m}}\in\set{0,-2},%
\end{cases}%
\label{eqn:edgelength_derivative}%
\end{align}
depends only on static geometric quantities, where the potential discontinuities of $\evaluate{A^{\prime\prime}_k\fof{\signdist}}{s=\projvertpos{i}}$ can be removed by different choices of the status assignment. %
\begin{figure}[htbp]%
\centering%
\subfigure[degenerate face intersection configurations; cf.~\reftab{intersection_status_edge}]{\label{fig:face_intersection_area_01}%
\includegraphics[page=1]{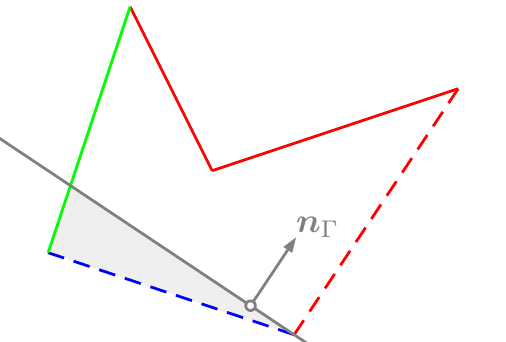}%
\includegraphics[page=2]{face_intersection}%
\includegraphics[page=3]{face_intersection}%
}%
\\%
\subfigure[normalized interior face area (\refeqnsave{intersected_face_area_reduced}) and derivatives (\refeqnsave{intersected_face_area_reduced_derivative}) as function of normalized position $\tilde{\signdist}$, where the dashed vertical lines correspond to the configurations in \subref{fig:face_intersection_area_01} (left to right)]{\label{fig:face_intersection_area_03}%
\includegraphics[page=1]{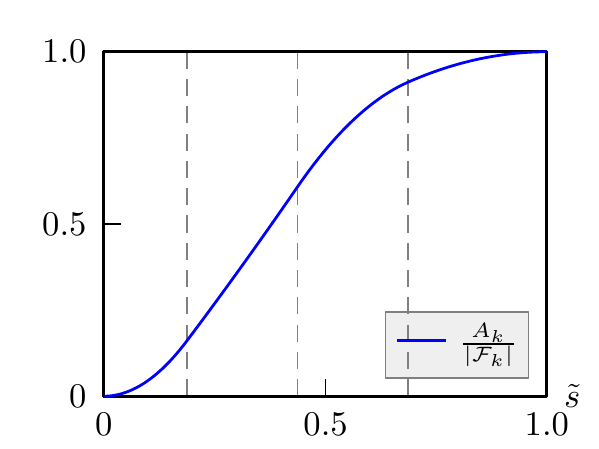}%
\includegraphics[page=2]{face_intersection_area}%
\includegraphics[page=3]{face_intersection_area}%
}%
\caption{Degenerate intersection configurations and corresponding discontinuities of second derivative of face area (\refeqnsave{intersected_face_area_reduced_derivative}) for polygonal face $\vfface_k$ with vertices $\{(1,3),(7,1),(11,7),(5,5),(3,9)\}$, base point $\xbase={[1,3]}\transpose$ and PLIC normal $\plicnormal=\frac{[4,6]\transpose}{\sqrt{52}}$.}%
\label{fig:face_intersection_area}%
\end{figure}
\refFig{face_intersection_area} illustrates the concept, whose main implications can be cast as follows: %
\begin{enumerate}
\item The assignment of the edge length derivative based on the edge status corresponds to choosing the left- (\refeqn{edgelength_derivative}; \textcolor{orange}{$\blacksquare$} in \reffig{face_intersection_area_03}) or right-sided limit, i.e. %
\begin{align}
A^{\prime\prime}_k\fof{\projvertpos{i}}\defeq\lim_{s\nearrow \projvertpos{i}}A^{\prime\prime}_k\fof{\signdist}%
\qquad\text{or}\qquad%
A^{\prime\prime}_k\fof{\projvertpos{i}}\defeq\lim_{s\searrow \projvertpos{i}}A^{\prime\prime}_k\fof{\signdist}.%
\end{align}%
In theory, each interior\footnote{The boundary discontinuities $\set{\projvertposmin,\projvertposmax}$ do not require to be removed for the problem under consideration here.} discontinuity $\projvertposmin<\projvertpos{i}<\projvertposmax$ may be removed by an individual linear combination of the respective left- and right-sided limit. However, within this work, we choose the left-sided limits; cf.~\refeqn{edgelength_derivative}. %
\item While, from a conceptual point of view, the continuation of $A^{\prime\prime}_k$ is important, the numerical results shown in section~\ref{sec:numerical_results} are barely affected by the choice of the status assignment, cf.~\refeqn{edgelength_derivative}. This is to be expected, since the probability of the iteration producing some $\signdist^n\in\set{\projvertpos{i}}$ is zero. 
\item For a degenerately intersected edge $\vfedge_{k,m}$, the assignment of $l_{k,m}$ (\refeqn{edgelength}) and $l^\prime_{k,m}$ (\refeqn{edgelength_derivative}) depends on the intersection status of its parenting face (index $k$).
\end{enumerate}
%
%
\begin{note}[Degenerately intersected faces ($\vfstatus{\vfface_k}=3$)]\label{note:edgeface_on_plic}%
The reasoning behind the removal of the discontinuities of $A^{\prime\prime}_k\fof{\signdist}$ analogously applies to $\polyvol^{\prime\prime}\fof{\signdist}$, i.e.\ the assignment of the status of topologically inferior entities (faces) corresponds to the left- or right-sided limits. As for the faces, we choose the left-sided limits to resolve the interior discontinuities $\projvertposmin<\projvertpos{i}<\projvertposmax$; cf.~\reffig{face_intersection_area}. %
\end{note}
%
%
\subsection{On the choice of the initial value $\signdist^0$}\label{subsec:initial_value}%
\begin{figure}[htbp]
\null\hfill%
\subfigure[deviation $\vofdeviation$ and cubic spline approximation $\vofdeviation_3$ for unit cube with normal $\plicnormal=\frac{{[1,-3,2]}\transpose}{\sqrt{14}}$ and $\refvof=\frac{27}{100}$.]{\includegraphics{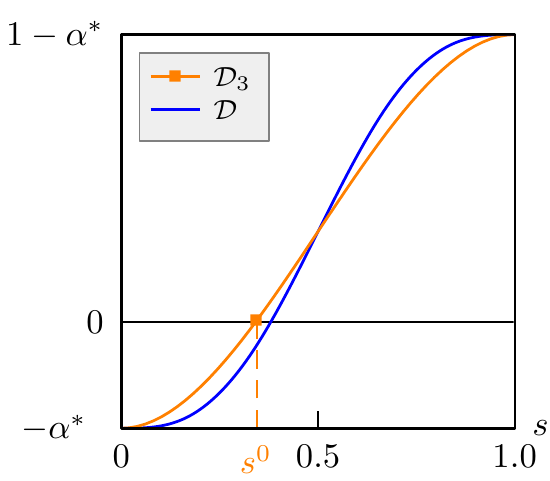}\label{fig:algorithm_initial_position_01}}%
\hfill%
\subfigure[Number of instances (in \%, rounded) where the PLIC normal $\plicnormal\in\protect\normalset{}$ (see~subsection~\ref{subsec:objective_conclusion} for details) and face normal $\vn_{\vfface,k}$ are aligned for faces $\vfface_k\subset\partial\mathcal{H}_{\protect\polyhedron*}$; cf.~\refnote{boundary_value_volume_derivative}.]{\includegraphics{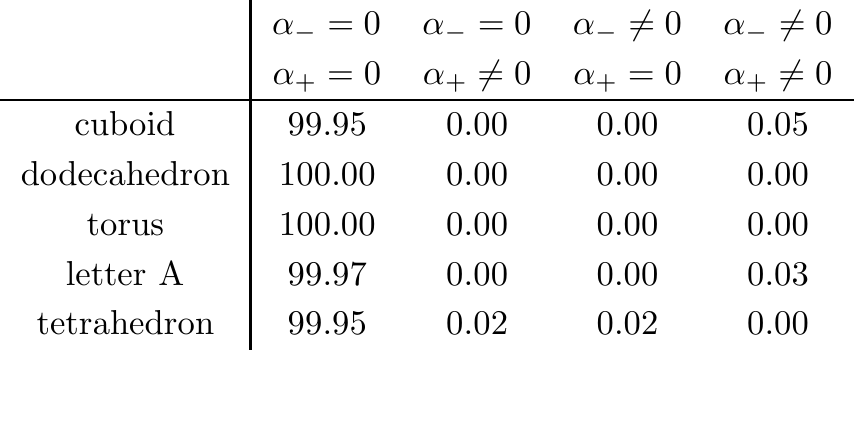}\label{fig:algorithm_initial_position_02}}%
\hfill\null%
\caption{Illustration of global cubic approximation (left; cf.~\refeqnsave{volume_fraction_cubic_approximation}) and number of instances in $\protect\normalset{}$ with non-trivial derivatives at the boundary (right).}%
\label{fig:algorithm_initial_position}%
\end{figure}

Numerical experiments have shown that, even for seemingly well-posed configurations of $\set{\polyhedron*,\plicnormal,\refvof}$, computing the initial value for \refeqn{position_minimum_newton_iteration} by linear interpolation, i.e.\ $\signdist^0\defeq\projvertposmin+\refvof\projvertsize$, may inhibit convergence to the desired root $\signdistref$. Hence, following \citet{XXX_2020_ivof}, the root of a global cubic approximation of the deviation $\vofdeviation$ (fulfilling the boundary conditions discussed in \refnote{boundary_value_volume_derivative}) provides a suitable initial value. Plugging the derivatives of the volume fraction at the lower and upper bound, say, $\polyvof^\prime_\pm\defeq\polyvof^\prime\fof{\projvertposminmax}$, into the standard cubic spline given in \refeqn{cubic_spline_interpolation} yields %
\begin{align}
\vofdeviation_3\fof{\tilde{\signdist};\refvof}&=%
\brackets*{\polyvof^\prime_-\projvertsize+\polyvof^\prime_+\projvertsize-2}\tilde{\signdist}^3+%
\brackets*{3-2\polyvof^\prime_-\projvertsize-\polyvof^\prime_+\projvertsize}\tilde{\signdist}^2+%
\polyvof^\prime_-\projvertsize\tilde{\signdist}-%
\refvof%
\quad\text{with}\quad%
\tilde{\signdist}=\frac{\signdist-\projvertposmin}{\projvertsize},
\label{eqn:volume_fraction_cubic_approximation}%
\end{align}
where \citet{XXX_2020_ivof} assumes $\polyvof^\prime_\pm\equiv0$ for reasons of computational efficiency; \reffig{algorithm_initial_position_01} illustrates the concept. %

Despite the fact that the results shown in \reffig{algorithm_initial_position_02} strongly depend on the choice of $\normalset{}$, this numercial observation supports the assumption of \citet{XXX_2020_ivof} ($\polyvof_\pm\equiv0$). While \citet{XXX_2020_ivof} employs a \textsc{Newton}-type method for the computation of $\signdist^0$ for reasons of numerical error control, we prefer to pursue an analytical approach: since the discriminant $\Delta\fof{\vofdeviation_3}=\nicefrac{27}{4}\refvof\brackets{1-\refvof}$ is positive for $0<\refvof<1$, $\vofdeviation_3$ exhibits three distinct real roots, one of which is located in $[\projvertposmin,\projvertposmax]$. From the formula of \textsc{Vi\`eta}, one obtains %
\begin{align}
s^0\fof{\refvof}\defeq\projvertposmin+\projvertsize\brackets*{\frac{1}{2}-\cos\brackets*{\frac{\arccos\fof{2\refvof-1}-2\pi}{3}}}.\label{eqn:initial_iteration_trivial}%
\end{align}
%
%
While it is theoretically possible to impose non-trivial \textsc{Neumann}-type boundary conditions for $\vofdeviation_3$ at $\projvertposmin$ and $\projvertposmax$, there are two major issues to consider: %
\begin{enumerate}
\item%
For general $\polyvof_\pm\neq0$, the evaluation of the analytic formulae for the roots becomes numerically challenging, due to the complexity of the involved expressions. \refFig{admissibility_rootclass} shows the sign of the discriminant $\Delta\fof{\vofdeviation_3}$ for $\refvof=\nicefrac{1}{2}$, which is decisive for the class of the roots and, hence, the respective analytical expression. In general, the evaluation of $\Delta\fof{\vofdeviation_3}$ requires to compute sixth powers of potentially very small numbers, which, in terms of numerical robustness, poses a highly complex task in itself. Their numerical computation, on the other hand, implies additional computational effort. Especially for $\refvof\approx0$ and $\refvof\approx1$, a robust computation requires careful implementation. %
\begin{figure}[htbp]%
\null\hfill%
\includegraphics[page=2]{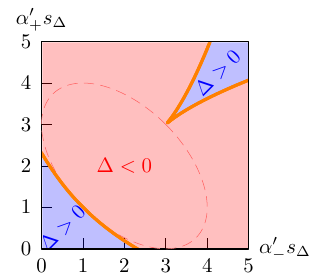}
\hfill\null%
\caption{Sign of discriminant $\Delta\fof{\vofdeviation_3}$ as a function of normalized boundary derivatives $\polyvof^\prime_\pm\projvertsize$.}%
\label{fig:admissibility_rootclass}%
\end{figure}%
\item There are configurations $\set{\polyvof^\prime_-,\polyvof^\prime_+}$ degrading the monotonicity of $\vofdeviation_3$, hence violating one of the properties of the (target) function, namely $\polyvof^\prime=\frac{\abs{\plicplane\cap\polyhedron*}}{\abs{\polyhedron*}}>0$. Computing the derivative of \refeqn{volume_fraction_cubic_approximation} yields a quadratic polynomial, whose roots correspond to the critical points of $\vofdeviation_3$. After some simple manipulations, one obtains %
\begin{align}
\set*{\polyvof^\prime_\pm\in[0,\infty)^2:%
3\brackets*{\polyvof^\prime_-+\polyvof^\prime_+-4\projvertsize^{-1}}^2+\brackets{\polyvof^\prime_--\polyvof^\prime_+}^2\leq12\projvertsize^{-2}%
\quad\text{or}\quad%
\brackets{\polyvof^\prime_-+\polyvof^\prime_+}\projvertsize\leq3}%
\label{eqn:vofdeviation_admissibility}%
\end{align}
as the set of admissible configurations, being a composition of an ellipse with semiaxes $\projvertsize^{-1}\brackets{\sqrt{2},\sqrt{6}}$ centered at $\projvertsize^{-1}\brackets{2,2}\transpose$ and rotated by $\frac{\pi}{4}$ and a triangle; cf.~\reffig{admissibility_montonicity}. %
\begin{figure}[htbp]%
\null\hfill%
\includegraphics[page=1]{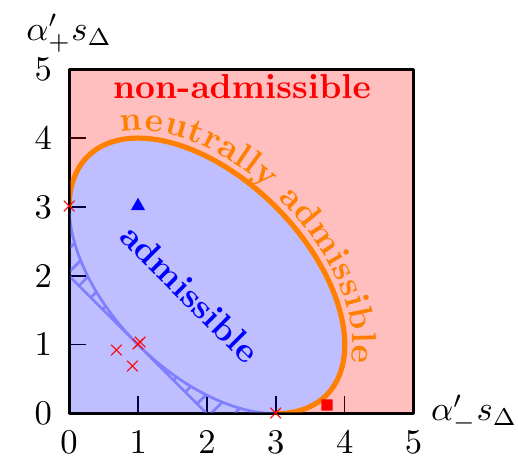}
\hfill%
\includegraphics[page=5]{hermite}
\hfill%
\includegraphics[page=6]{hermite}
\hfill\null%
\caption{Admissibility chart of normalized boundary derivatives $\polyvof^\prime_\pm\projvertsize$ with illustrations of admissible (\textcolor{blue}{$\blacktriangle$}, center) and non-admissible (\textcolor{red}{$\blacksquare$}, right) approximations $\vofdeviation_3$.}%
\label{fig:admissibility_montonicity}%
\end{figure}
Despite that none of the numerical test cases presented in section~\ref{sec:numerical_results} (indicated by \textcolor{red}{$\times$}) suffers from being non-admissible, the admissibility requires to be probed in any robust implementation. %
\end{enumerate}
Furthermore, numerical experiments have shown that the results presented in section~\ref{sec:numerical_results} do not benefit from considering $\polyvof_\pm\neq0$, once more supporting the assumption of \citet{XXX_2020_ivof} discussed above. Thus, in what follows, we employ \refeqn{initial_iteration_trivial} to obtain the initial iteration. %
%
%
\subsection{On the computation of the step size $\Delta\signdist$}\label{subsec:stepsize}%
The analytic expression of the volume fraction obtained in \refeqn{polyhedron_volume_signed_distance} implies that its derivatives with respect to the signed distance $\signdist$ can be easily evaluated, suggesting an exploitation within a higher-order rootfinding algorithm. Thus, the present work introduces an algorithm based on \textbf{implicit bracketing} with \textbf{locally quadratic} approximation. The upcoming subsections provide the key ideas underlying the numerical algorithm proposed in section~\ref{sec:implementation}. %
%
%
\subsubsection{Implicit bracketing}\label{subsubsec:implicit_bracketing}%
\begin{figure}[htbp]%
\null\hfill%
\includegraphics[page=1]{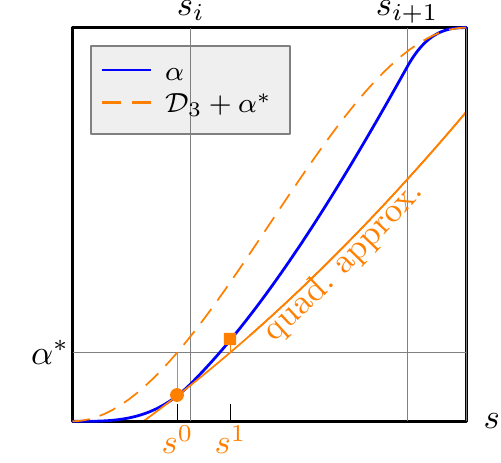}
\hfill%
\includegraphics[page=2]{implicit_bracketing}
\hfill%
\includegraphics[page=3]{implicit_bracketing}
\hfill\null%
\caption{Implicit bracketing and locally quadratic approximation of volume fraction $\polyvof\fof{\signdist}$; cf.~subsection~\ref{subsubsec:taylor_step}.}%
\label{fig:implicit_bracketing_illustration}%
\end{figure}

Recall from subsection~\ref{subsec:volume_computation} that, within a bracket $\mathcal{B}_i\defeq[\signdist_i,\signdist_{i+1}]$, the volume fraction $\polyvof\fof{\signdist}$ is an increasing cubic polynomial, denoted by $\mathcal{S}_i\fof{\signdist}$. For any given $\signdist^n\in\mathcal{B}_i$, the polynomial reads %
\begin{align}
\mathcal{S}_i\fof{z;\signdist^n}=%
\frac{\polyvof^{\prime\prime\prime}\fof{\signdist^n}}{6}\brackets*{z-\signdist^n}^3+%
\frac{\polyvof^{\prime\prime}\fof{\signdist^n}}{2}\brackets*{z-\signdist^n}^2+%
\polyvof^{\prime}\fof{\signdist^n}\brackets*{z-\signdist^n}+%
\polyvof\fof{\signdist^n}.\label{eqn:third_order_approximation}%
\end{align}%
Hence, the truncation of the polyhedron $\polyhedron*$ at any $\signdist^n$ (implicitly) provides the full information of $\polyvof\fof{\signdist}$ within the containing bracket $\mathcal{B}_i$. Exploiting that $\polyvof_i=\mathcal{S}_i\fof{\signdist_i;\signdist^n}$ and $\polyvof_{i+1}=\mathcal{S}_i\fof{\signdist_{i+1};\signdist^n}$ suggests the following strategy: %
\begin{enumerate}%
\item if the current iteration $\signdist^n$ is \textit{not} located in the target bracket $\refbracket$ ($\refvof<\polyvof_i$ or $\polyvof_{i+1}<\refvof$), the next iteration is obtained from locally quadratic approximation (\reffig{implicit_bracketing_illustration}, left). %
\item if the current iteration $\signdist^n$ is located in the target bracket $\refbracket$ ($\polyvof_i\leq\refvof\leq\polyvof_{i+1}$), the sought $\signdistref$ corresponds to the root of $\mathcal{S}_i-\refvof$, requiring no further truncation (\reffig{implicit_bracketing_illustration}, center). %
\end{enumerate}%
\refFig{implicit_bracketing_illustration} illustrates the components of the strategy outlined above: with $\signdist^0\not\in\mathcal{B}_i$ (left), a quadratic approximation yields $\signdist^{1}$, which lies within the target bracket $\refbracket=\mathcal{B}_i$ (center). The rightmost configuration already starts with $\signdist^0\in\refbracket$, such that the spline interpolation directly yields the sought position $\signdistref$, implying that only a single truncation is required. As we shall see in section~\ref{sec:numerical_results}, configurations of this type are often obtained for $\refvof\approx0$ and $\refvof\approx1$. %
\begin{note}[Comparison to \citet{XXX_2020_ivof}]\label{note:ccs_comparison}%
\refFig{ccs_step_illustration} shows that the CCS method of \citet{XXX_2020_ivof} also allows to obtain an exact interpolation of the volume fraction $\polyvof$. However, there are two major differences to the present algorithm: %
\begin{enumerate}
\item In general, the CCS method requires two subsequent iterations within the same bracket, i.e.\ $\set{\signdist^{n-1},\signdist^n}\subset[\signdist_i,\signdist_{i+1}]$ (assuming that $\signdist^n>\signdist^{n-1}$ without loss of generality). 
\item If two iterations $\set{\signdist^{n-1},\signdist^n}\subset[\signdist_i,\signdist_{i+1}]$ are found, the CCS method only exploits the exact interpolation in $[\signdist^{n-1},\signdist^n]$. Assume that $\signdistref\in[\signdist_i,\signdist_{i+1}]\setminus[\signdist^{n-1},\signdist^n]$, i.e.\ the spline interpolation governs the sought $\signdistref$ without a sign change between the iterations, i.e.\ $\vofdeviation\fof{\signdist^{n-1}}\vofdeviation\fof{\signdist^{n}}>0$. Then, a linear approximation is applied to obtain $\signdist^{n+1}$, implying another truncation, despite the fact that the target bracket $\refbracket=[\signdist_i,\signdist_{i+1}]$ was already found. %
\end{enumerate}
As will be shown in section~\ref{sec:numerical_results}, these properties, in combination with a sufficiently small tolerance $\zerotol$, yield an average margin of one polyhedron truncation for the proposed method in comparison to CCS. %
\end{note}
\begin{note}[Comparison to \citet{JCP_2016_anvc}]\label{note:cibrave_comparison}%
While the present work resorts to the same basic building blocks as CIBRAVE, the novel method of volume computation introduced in subsection~\ref{subsec:volume_computation} allows for considerable improvement and a local formulation. The main differences of the algorithm of \reffig{main_algorithm_flowchart} and CIBRAVE are:\\[10pt]%
\null\hfill%
\begin{tabular}{c||c|c}
&\textbf{CIBRAVE}&\textbf{present work}\\%
\hline%
\textbf{initial iteration $\signdist^0$}&linear interpolation&cubic interpolation (\refeqn{volume_fraction_cubic_approximation})\\%
\hline%
\textbf{analytical representation}&additional computation&by-product of truncation\\%
\textbf{of $\polyvof\fof{\signdist}$}&(\textbf{only} at $\set{\projvertpos{i}}$)&(at \textbf{any} $\signdist^n$)\\%
\hline%
\textbf{update $\signdist^{n+1}$}&linear interpolation&quadratic approximation\\%
&based on $\set{\polyvof\fof{\projvertpos{i}}}$&around $\signdist^n$ (see~\ref{subsubsec:taylor_step})\\%
\multicolumn{3}{c}{}%
\end{tabular}%
\hfill\null\\%
Let us note in passing that CIBRAVE requires topological connectivity on a cell level, whereas the present work only requires a list of faces. %
\end{note}
%
%
\subsubsection{Locally quadratic approximation (Taylor)}\label{subsubsec:taylor_step}%
\begin{figure}[htbp]%
\null\hfill%
\includegraphics[page=1]{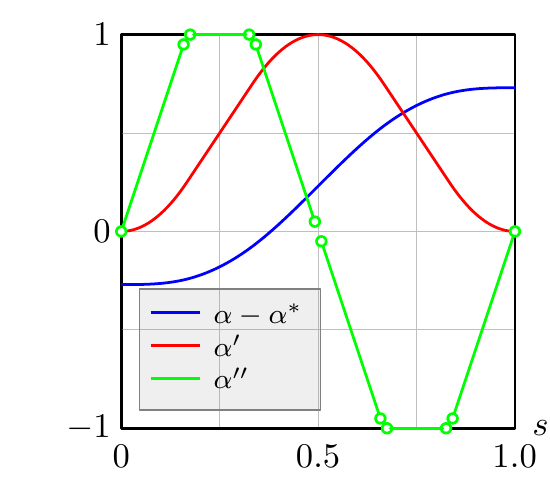}%
\hfill%
\includegraphics[page=2]{algorithm_example_steps}%
\hfill\null%
\caption{Volume fraction deviation $\vofdeviation$ and derivatives for a cuboid with $\plicnormal=\frac{{[1,-3,2]}\transpose}{\sqrt{14}}$ and $\refvof=\frac{27}{100}$ (left) and associated \textsc{Newton} (\refeqnsave{newton_step}) and \textsc{Taylor} (\refeqnsave{taylor_step}) steps (right).}%
\label{fig:newton_taylor_step_illustration}%
\end{figure}

\begin{figure}[htbp]
\null\hfill%
\includegraphics[page=2]{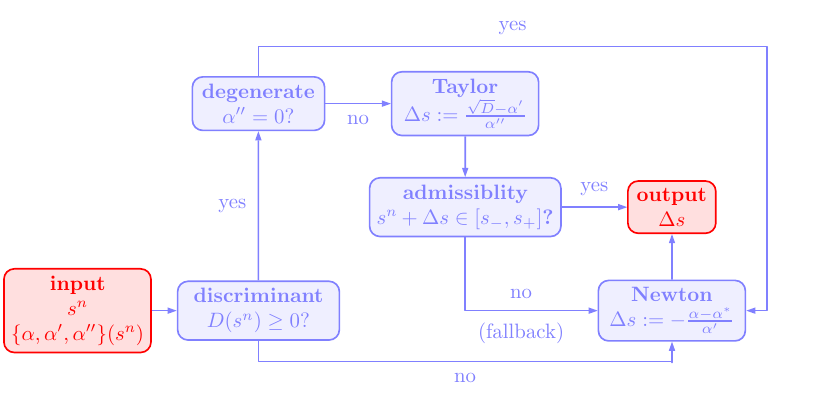}
\hfill\null%
\caption{Flowchart of \textsc{Taylor} step computation.}%
\label{fig:taylorstep_flowchart}%
\end{figure}

Locally, at some given $\signdist^n$, the deviation $\vofdeviation$ can be approximated by %
\begin{align}
T_{\vofdeviation}\fof{z;\signdist^n}=\frac{\polyvof^{\prime\prime}\fof{\signdist^n}}{2}\brackets{z-\signdist^n}^2+\polyvof^{\prime}\fof{\signdist^n}\brackets{z-\signdist^n}+\polyvof\fof{\signdist^n}-\refvof,%
\end{align}
from whose roots (assuming, for the moment, that $\polyvof^{\prime\prime}\fof{\signdist^n}\neq0$) we obtain %
\begin{align}
\brackets[s]{\Delta\signdist_{\mathrm{Taylor}}}\fof{\signdist^n}=\brackets*[s]{\frac{\sqrt{D}-\polyvof^{\prime}}{\polyvof^{\prime\prime}}}\fof{\signdist^n}%
\quad\text{with the discriminant}\quad%
D=\brackets*[s]{\polyvof^{\prime}\polyvof^{\prime}-2\brackets{\polyvof-\refvof}\polyvof^{\prime\prime}}\fof{\signdist^n}.\label{eqn:taylor_step}%
\end{align}
Exploiting $\polyvof\fof{\signdistref}=\refvof$ and enforcing $\signdistref$ to be a fixed point of the iteration, i.e.\ $\Delta\signdist\fof{\signdistref}\stackrel{!}{=}0$, eliminates the quadratic ambiguity, i.e.\ $\frac{-\sqrt{D}-\refvof}{\polyvof^{\prime\prime}}$ is ruled out in favor of the expression given in \refeqn{taylor_step}. %
For vanishing second derivative, the \textsc{Taylor} step degenerates to the one obtained by linear approximation, i.e. %
\begin{align}
\lim\limits_{\polyvof^{\prime\prime}\to0}\brackets*[s]{\Delta\signdist_{\mathrm{Taylor}}}\fof{\signdist^n}=\brackets*[s]{-\frac{\polyvof-\refvof}{\polyvof^{\prime}}}\fof{\signdist^n}=\brackets*[s]{\Delta\signdist_{\mathrm{Newton}}}\fof{\signdist^n}.\label{eqn:newton_step}%
\end{align}
\refFig{newton_taylor_step_illustration} provides an illustration for a cuboid, where the flowchart in \reffig{taylorstep_flowchart} gathers the details of the numerical implementation. %

\begin{note}[Locally third-order approximation]\label{note:locally_third_order_approximation}%
While it is theoretically possible to apply a third-order approximation, we restrain from doing so for two reasons:%
\begin{enumerate}
\item Numerical experiments have shown that, in general, $\abs{\polyvof\fof{\signdist}-\mathcal{S}_i\fof{\signdist;\signdist^n}}$ becomes prohibitively large for $\signdist\not\in\mathcal{B}_i$, being attributable to the insufficient regularity of $\polyvof\fof{\signdist}$ (right); cf.~\refnote{boundary_value_volume_derivative}. This disadvantageous effect prevails prominently in the beginning of the iteration. %
\item In cases where the third order approximation proves sufficiently accurate, the second order approximation provides similiar results, however without the requirement for numerical computation of the root (left); cf.~subsection~\ref{subsubsec:taylor_step} below. %
\end{enumerate}
\null\hfill%
\includegraphics[page=1]{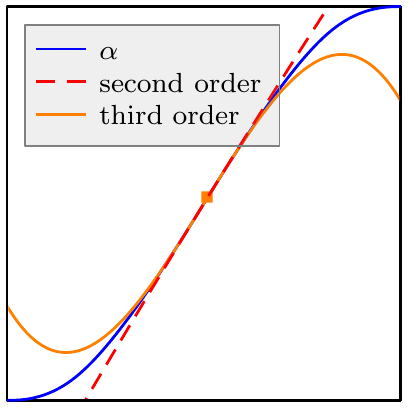}
\hfill%
\includegraphics[page=2]{vof_approx}
\hfill\null%
\end{note}%
%
%
%
%
\section{Implementation}\label{sec:implementation}%
\begin{figure}[htbp]%
\null\hfill%
\includegraphics[page=1]{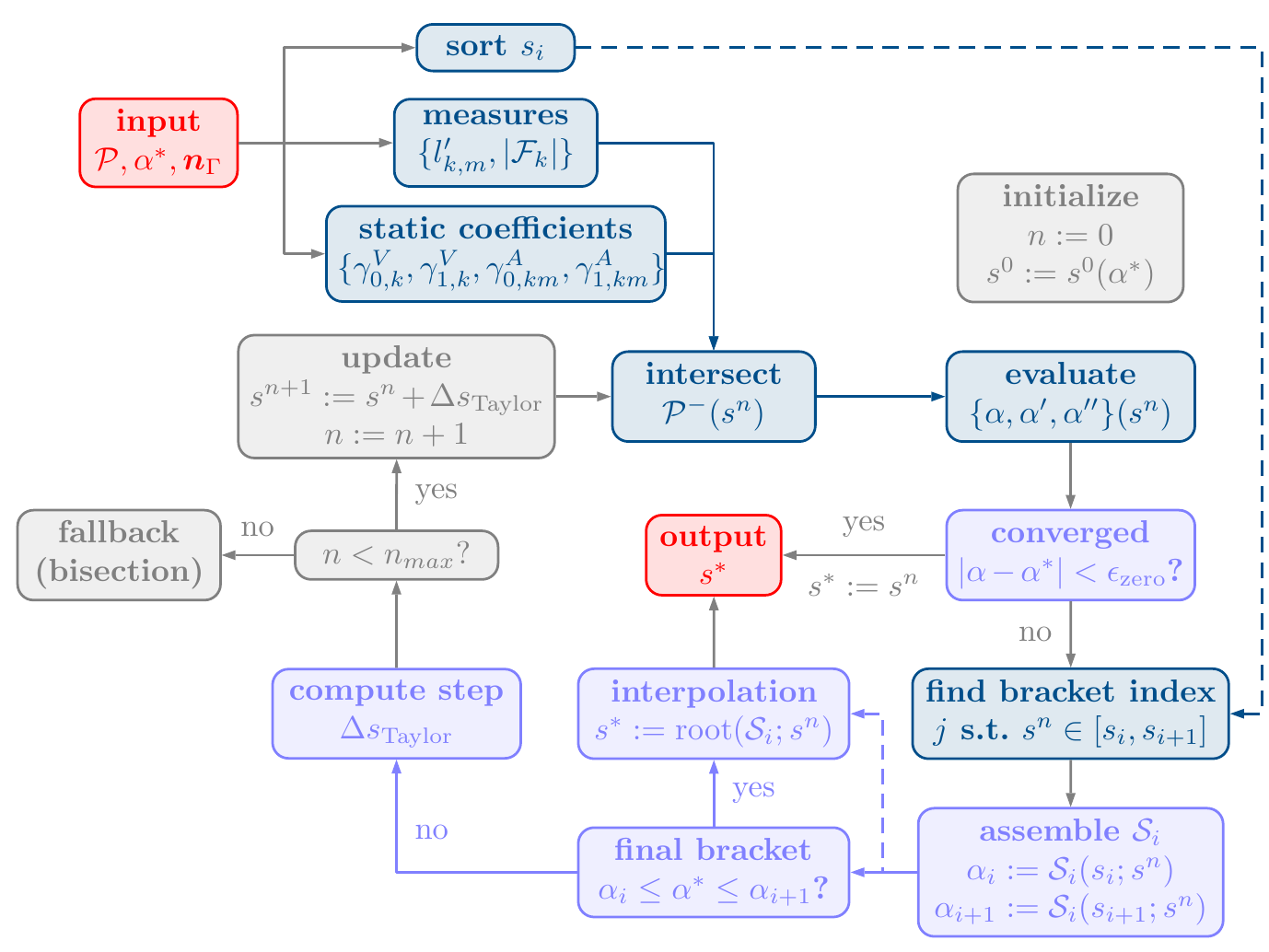}%
\hfill\null%
\caption{Flowchart of the main algorithm. For all numerical testcases reported here, a tolerance of $\zerotol=\num{e-12}$ was used.}%
\label{fig:main_algorithm_flowchart}%
\end{figure}

\refFig{main_algorithm_flowchart} provides a flowchart of the algorithm. After reading the input parameters $\set{\polyhedron*,\refvof,\plicnormal}$, representing the polyhedron, volume fraction and normal, respectively, the static measures and coefficients (cf.~\refeqs{volume_coefficients}, \refeqno{face_coefficients}, \refeqno{edgelength}-\refeqno{edgelength_derivative}) are computed and the signed distances of the vertices $\signdist_i$ are sorted in ascending order (see~section~\ref{sec:mathematical_formulation} for details). Starting with the initial value $\signdist^0$ given in subsection~\ref{subsec:initial_value}, within each iteration (superscript $n$), the truncation of the polyhedron yields the volume fraction and its derivatives. From the index $i$ of the bracket containing the current iteration $\signdist^n$, the governing cubic polynomial $\mathcal{S}_i$ is assembled, allowing to evaluate the interval $[\polyvof_i,\polyvof_{i+1}]$. For $\refvof\in[\polyvof_i,\polyvof_{i+1}]$, implying that $\signdist^n\in\refbracket$, the sought position $\signdistref$ corresponds to the root of the cubic polynomial $\mathcal{S}_i$, which is obtained numerically resorting to the algorithm described in appendix~\ref{app:rootfinding_spline}. In the case where $\mathcal{B}_i$ does not contain the sought position $\signdistref$, the locally quadratic approximation of subsubsection~\ref{subsubsec:taylor_step} provides the next iteration $s^{n+1}$. By design, the quadratic approximation may produce an update $\signdist^{n+1}$ located in a different bracket than the current iteration $\signdist^n$. Because the admissibility ($\signdist^{n+1}\stackrel{!}{\in}[\projvertposmin,\projvertposmax]$) cannot be enforced a priori, an appropriate check is performed, resorting to a \textsc{Newton} step\footnote{In theory, the \textsc{Newton} step may also fail to produce an admissible update. However, we have not encountered such a configuration.} if $\signdist^{n+1}\not\in[\projvertposmin,\projvertposmax]$; cf.~\reffig{taylorstep_flowchart}. %
It is possible that, after an update with $\Delta\signdist_{\mathrm{Taylor}}$, the deviation $\vofdeviation\fof{\signdist^{n+1}}$ becomes smaller than $\zerotolerance$. These cases emerge for a plane $\plicplane$ being \textit{almost} parallel to one of the faces $\vfface_k$ for $\refvof\approx0$ or $\refvof\approx1$. This combination yields a clustered set $\set{\signdist_i}$ of signed distances of the respective face vertices, implying a very small size of the target bracket $\refbracket$. In these cases, a quadratic approximation around some $\signdist^n\not\in\refbracket$, but in the vicinity of $\refbracket$, is likely to be sufficiently accurate. However, their incidence barely reaches statistical significance. %
Should the iteration not yield the sought $\signdistref$ with the prescribed number of iterations $n_\mathrm{max}=\num{100}$, the algorithm resorts to a bisection method. %

\begin{remark}[Implementation complexity]%
As can be seen from the flowchart in \reffig{main_algorithm_flowchart}, the algorithm presented in this paper heavily relies on the volume computation descibed in \ref{subsec:volume_computation}. Despite its face-based formulation, an integration into an existing legacy code, such as, e.g., openFOAM, requires some implementation effort. While the presented method gets along with fairly basic geometric operations, the CIBRAVE algorithm of \citet{JCP_2016_anvc} moreover requires establishing the topological connectivity of truncated polyhedral cells, which induces significant implementation complexity. In contrast, the CCS method of \citet{XXX_2020_ivof} only resorts to the derivative of $\polyvof^\prime$, which is proportinal to $\abs{\plicplane\cap\polyhedron*}$. For all methods, the implementation complexity of the rootfinding routines is negligible in comparion the geometric operations. 
\end{remark}%

\begin{impldetail}[Object orientation]%
The algorithm presented in this paper was implemented in Fortran 2008 (8-byte real/4-byte integer), employing object-orientation wherever possible. In \reffig{main_algorithm_flowchart}, light blue boxes belong to the root-finding part and the dark blue boxes correspond to methods of the abstract class \textsf{polyhedron}; cf.~subsection~\ref{subsec:notation}. Since the computation of the measures and coefficients (\refeqs{volume_coefficients}, \refeqno{face_coefficients}, \refeqno{edgelength}-\refeqno{edgelength_derivative}) can be implemented in a generic way, the concretization of the derived classes (\textsf{cuboid} etc., cf.~\reftab{polyhedron_types}) only requires to implement the polyhedron definition, i.e.\ the assignment of vertices and topology. %
\end{impldetail}
%
%
%
%
\section{Numerical results}\label{sec:numerical_results}%
\begin{figure}[htbp]
\null\hfill%
\subfigure[dodecahedron]{\includegraphics[width=6cm]{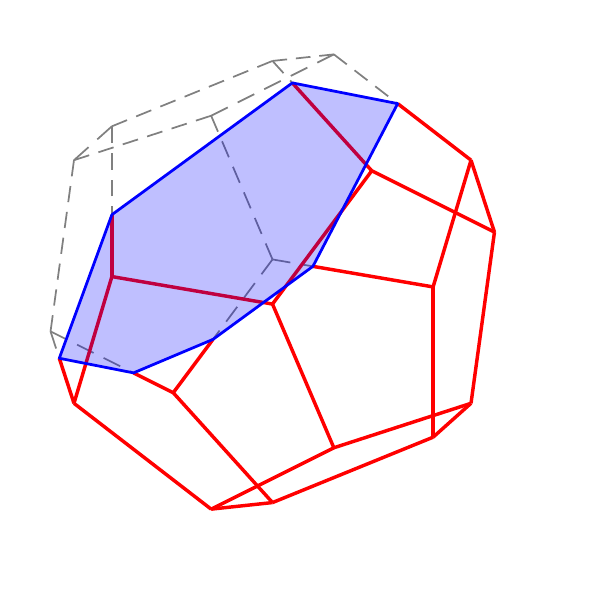}\hfill\includegraphics[page=1]{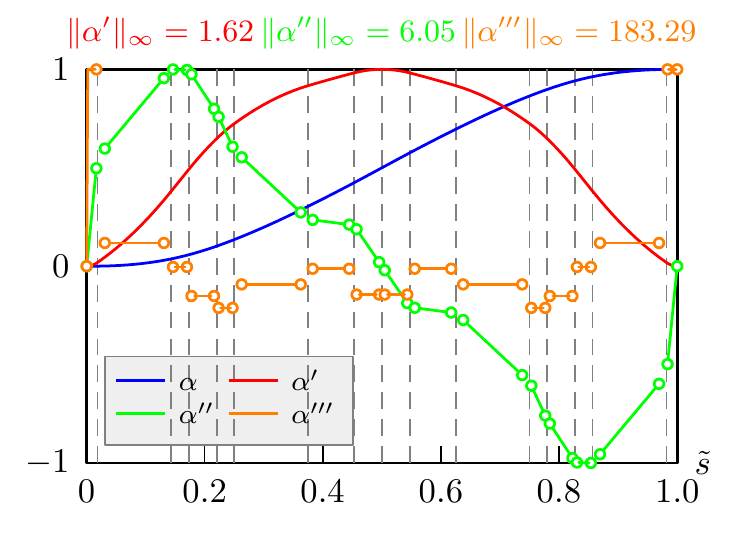}\label{fig:illustration_vof_01}}%
\hfill\null%
\\%
\null\hfill%
\subfigure[torus]{\includegraphics[width=6cm]{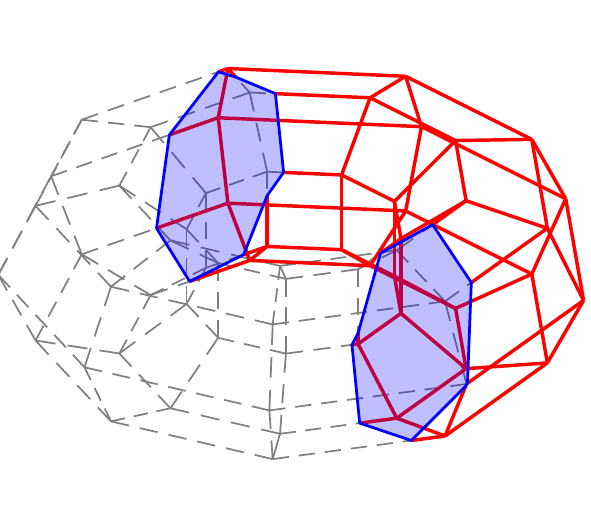}\hfill\includegraphics[page=1]{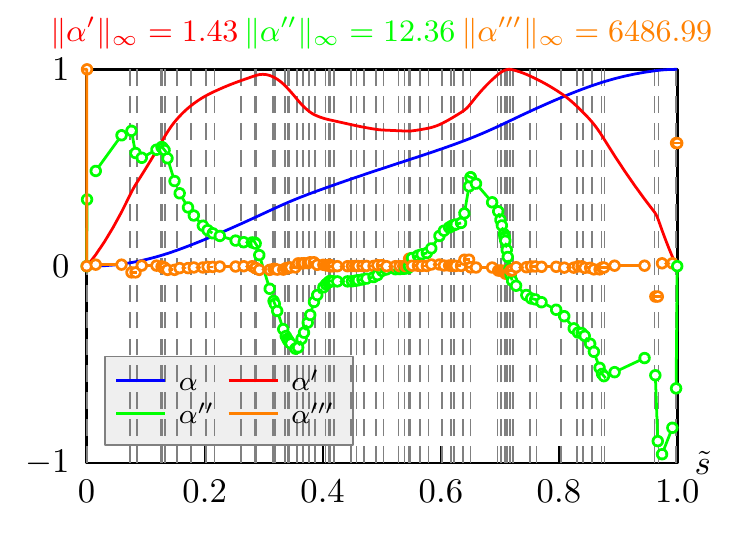}\label{fig:illustration_vof_02}}%
\hfill\null%
\\%
\null\hfill%
\subfigure[letterA]{\includegraphics[width=6cm]{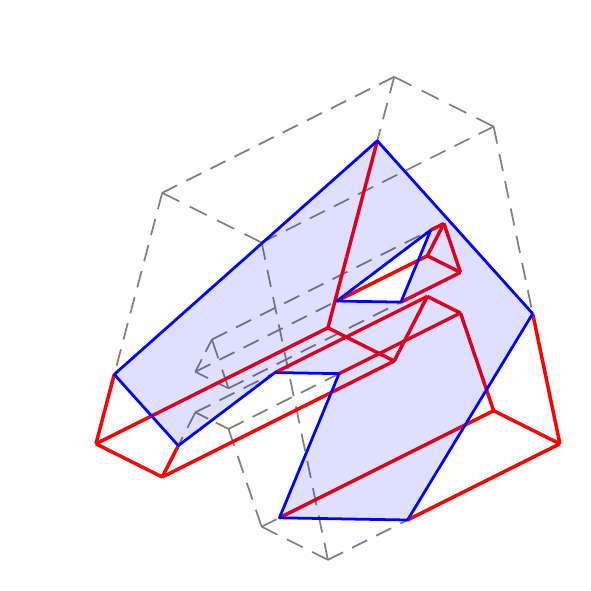}\hfill\includegraphics[page=1]{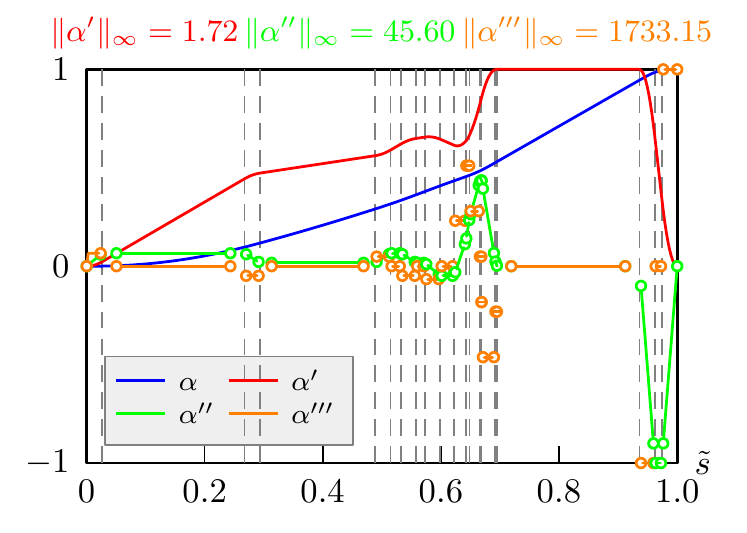}\label{fig:illustration_vof_03}}%
\hfill\null%
\caption{Illustration of intersected polyhedra with volume fraction as function of normalized parameter $\tilde{\signdist}$. The dashed vertical lines indicate the bracket boundaries $\signdist_i$, where the circles indicate the potential discontinuity at $\projvertpos{i}$. For the reader's convenience, the derivatives are scaled by the respective $\protect\norm{\cdot}_\infty$.}%
\label{fig:illustration_vof_polyhedra}%
\end{figure}

In what follows, we assess the algorithm presented in section~\ref{sec:implementation} for a comprehensive set of instances $\set{\polyhedron*,\refvof,\plicnormal}$, where \reftab{polyhedron_types} gathers the classes of polyhedra $\polyhedron*$ and subsection~\ref{subsec:objective_conclusion} defines the set of volume fractions $\refvof\in\vofset{}$ and normals $\plicnormal\in\normalset{}$. Separate processing and aggregation of the results highlights the influence of $\refvof$ and $\plicnormal$ in subsections~\ref{subsec:influence_vof} and \ref{subsec:influence_normal}, respectively. %
Before discussing the numerical experiments in detail, however, it is instructive to illustrate the volume fraction and its derivatives as a function of the plane position $\signdist$ for some selected instances in \reffig{illustration_vof_polyhedra}. The strong variance of the third derivatives in \reffig{illustration_vof_03} vividly illustrates the argument of \refnote{locally_third_order_approximation}, stating that, in general, a third order approximation $\mathcal{S}_i\fof{\signdist;\signdist^n}$ around some $\signdist^n\in\mathcal{B}_i$ produces large deviations $\abs{\polyvof\fof{\signdist}-\mathcal{S}_i\fof{\signdist;\signdist^n}}$ outside the bracket containing $\signdist^n$, i.e.\ for $\signdist\not\in\mathcal{B}_i$; cf.~\refeqn{third_order_approximation}. %
\paragraph{A suitable and robust measure for performance assessment.} While total absolute execution times constitute an intuitive measure for the computational performance of an algorithm, the problem at hand intrinsically features a more suitable indicator: most of the literature contributions discussed in subsection~\ref{subsec:literature_review} state that the polyhedron truncation accounts for the largest share in computation time; see, e.g., \citet[fig.~16]{JCP_2016_anvc} and \citet[fig.~8]{XXX_2020_ivof}. This property becomes especially prominent for non-convex polyhedra, where the truncation may produce disconnected sub-polyhedra, whose convex decomposition poses a challenging and computationally expensive task on its own. Despite the fact that the method proposed in subsection~\ref{subsec:volume_computation} significantly reduces the computational effort by avoiding the usage of topological connectivity, the inherent complexity of the geometric operations remains predominant. Thus, we assess the performance in terms of the numbers of truncation operations\footnote{For the algorithm given in \reffig{main_algorithm_flowchart}, the number of truncations corresponds to the number of iterations, i.e.\ $N_{\mathrm{trunc}}=N_{\mathrm{iter}}$.} per instance, i.e.\ $N_{\mathrm{trunc}}=N_{\mathrm{trunc}}\fof{\polyhedron*,\refvof,\plicnormal}$, corresponding to the final value of $n$ in the flowchart of \reffig{main_algorithm_flowchart}. Furthermore, we employ the execution times $t\fof{\polyhedron*,\refvof,\plicnormal}$ to establish a qualitative idea of the relative differences between the classes of polyhedra. The execution time\footnote{processor: Intel Core i7-7700HQ, compiler: gfortran 8.1.0 (with optimization \texttt{-O3}).} is obtained by subtracting the results of \texttt{MPI\_WTIME()}, called respectively at the begin and end of the subroutine implementing \reffig{main_algorithm_flowchart}, i.e.\ it includes all secondary operations, e.g., sorting $\set{\projvertpos{i}}$. %
\paragraph{Comparison to literature.}%
Subsection~\ref{subsec:objective_conclusion} and \reftab{polyhedron_types} outline the structure of numerical experiments conducted within this section and define the sample set for the volume fraction $\refvof$ and normal orientation $\plicnormal$. 
Out of the literature contributions discussed in subsection~\ref{subsec:literature_review}, only the manuscripts of \citet{JCP_2016_anvc} and \citet{XXX_2020_ivof} contain disaggregated data suitable for a meaningful comparison, where only \citet{XXX_2020_ivof} contains data for non-convex polyhedra. For example, \citet{JCP_2019_ncaa} only report \textit{relative} CPU times, normalized by the average runtime for a quadrangular cell. Also, they do not comment on the coverage of the measurements in terms of algorithmic elements. Thus, based on the data published in their manuscript, no sensible comparison can be made. %
Furthermore, \citet{JCP_2016_anvc} employ randomly distributed normals $\plicnormal$, while \citet{XXX_2020_ivof} resorts to the sample set given in subsection~\ref{subsec:objective_conclusion}. %
%
%
%
%
%
\subsection{Influence of the volume fraction $\refvof$}\label{subsec:influence_vof}%
\refFig{performance_statistics} gathers the aggregated results, where the figures in the respective rows share the following properties: %
{\renewcommand{\theenumi}{\roman{enumi}}
\begin{enumerate}
%
%
\item The volume fractions $\refvof$ on the abscissa are subject to a logarithmic stretch in the vicinity of zero and one, respectively. Between $\num{e-4}$ and $1-\num{e-4}$, the scaling remains linear. %
%
%
\item The averages are defined as %
\begin{align}
N_{\mathrm{av},\polyvof}\fof{\polyhedron*,\refvof}\defeq\frac{1}{N_{\plicnormal}}\sum\limits_{k=1}^{N_{\plicnormal}}{N_{\mathrm{trunc}}\fof{\polyhedron*,\refvof,\vn_{\plicplane,k}}}\quad\text{and}\quad%
N_{\mathrm{av}}\fof{\polyhedron*}\defeq\frac{1}{N_{\refvof}}\sum\limits_{m=1}^{N_{\refvof}}{N_{\mathrm{av},\polyvof}\fof{\polyhedron*,\refvof_m}},\label{eqn:result_average_definition}%
\end{align}
where $N_{\plicnormal}=\operatorname{size}\fof{\normalset{}}=2M_{\plicnormal}\brackets{M_{\plicnormal}+1}$ and $N_{\refvof}=\operatorname{size}\fof{\vofset{}}=M_{\refvof}+10$; cf.~subsection~\ref{subsec:objective_conclusion} for details. The execution times $t_{\mathrm{av},\polyvof}$ and $t_{\mathrm{av}}$ are defined analogously. For the reader's convenience, the left column contains their normalized counterpart, with the absolute scale $t_{\mathrm{av}}$ provided at the top of the respective figure. %
%
%
\item The solid dots ($\bullet$) represent the performance measures, given in \refeqn{result_average_definition}, obtained by the proposed algorithm, where the shaded regions indicate a tube with a width of twice the mean standard deviation. If available, the squares ($\blacksquare$) and triangles ($\blacktriangle$) show the results of \citet[fig.~17(d+e)]{JCP_2016_anvc} and \citet{XXX_2020_ivof}, respectively, for comparison. \citet{JCP_2016_anvc} obtain the relative execution times (last row in \reftab{performance_statistics}) from \num{e6} random combinations of $\refvof$ and $\plicnormal$. 
\end{enumerate}}%
The main implications of \reffig{performance_statistics} can be cast as follows: %
\begin{enumerate}
\item For all polyhedra under consideration, the average number of truncations is roughly symmetric to $\refvof=\nicefrac{1}{2}$, reflecting the symmetry of the positioning problem, since $\set{\polyhedron*,\refvof,\plicnormal}$ corresponds to $\set{\polyhedron*,1-\refvof,-\plicnormal}$. In general, however, the inverted order of $\set{\projvertpos{i}}$ spoils a corresponding symmetry of the initial iteration $\signdist^0$, obtained from the cubic interpolation $\vofdeviation_3$ given in \refeqn{volume_fraction_cubic_approximation}. %
\item The average number of truncations differs only slighty for the different polyhedra, which indicates a robust performance of the algorithm. The same holds for the relative execution times (left column in \reffig{performance_statistics}).%
\item For the cuboid, tetrahedron, dodecahedron and torus, there are no instances for which the fallback strategy has to be applied. For the letter A, out of all instances in subsection~\ref{subsec:objective_conclusion}, there are 8 (approximately \num{0.63}\textperthousand) fall-back cases. \refFig{letterA_taylor_newton_comparison} highlights the reasons: in essence, the applied second-order approximation bounces back and forth in a loop between two positions. Configurations of this type cannot be excluded a priori. 
However, this fallback-quota can safely be considered acceptable. %
\begin{figure}[htbp]
\null\hfill%
\subfigure[\textbf{no} convergence]{\includegraphics[page=1]{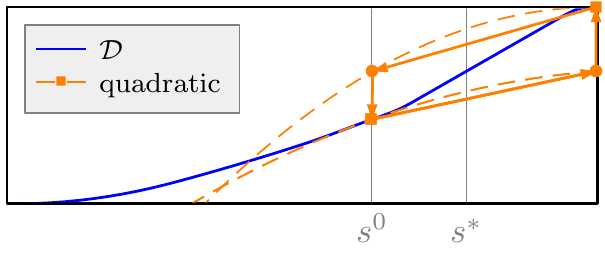}\label{fig:letterA_taylor_newton_comparison_02}}%
\hfill%
\subfigure[convergence after 3 iterations]{\includegraphics[page=2]{letterA_vof_fail}\label{fig:letterA_taylor_newton_comparison_04}}%
\hfill\null%
\caption{Comparison of convergence using \subref{fig:letterA_taylor_newton_comparison_02} \textsc{Taylor} and \subref{fig:letterA_taylor_newton_comparison_04} \textsc{Newton} step for polyhedron letter A: while the \textsc{Newton} method converges to the sought root within 3 iterations, the quadratic approximation bounces back and forth without convergence (indicated by \textcolor{orange}{$\blacksquare$}). In fact, all fallback cases reported for letter A are of this type.}%
\label{fig:letterA_taylor_newton_comparison}%
\end{figure}
\item For all polyhedra under consideration here, \reftab{performance_statistics} reports between \num{1.13} and \num{1.52} truncations to find $\signdistref$, where the relative execution times clearly reflect the geometric complexity of the polyhedra. %
\begin{table}[htbp]%
\caption{Aggregated execution statistics for proposed method; cf.~\reffig{performance_statistics} and table~3 in \protect\citet[p.~353]{JCP_2016_anvc}.}%
\label{tab:performance_statistics}%
\centering%
\begin{tabular}{c|ccccc}%
\multicolumn{6}{c}{}\\%
&\textbf{cuboid}%
&\textbf{dodecahedron}%
&\textbf{torus}%
&\textbf{letterA}%
&\textbf{tetrahedron}%
\\%
\hline%
$N_{\mathrm{av}}$ &\num{ 1.13}&\num{ 1.18}&\num{ 1.52}&\num{ 1.46}&\num{ 1.14}\\%
\hline%
$t_{\mathrm{av}}$ (\si{\micro\second})&\num{  2.5175}&\num{  5.7161}&\num{ 29.1500}&\num{  7.1372}&\num{  1.5967}\\%
relative&\num{  1.00}&\num{  2.27}&\num{ 11.58}&\num{  2.84}&\num{  0.63}\\%
relative (CIBRAVE)&1.00&\unifytime{18.39}{7.54}&--&--&\unifytime{4.61}{7.54}\\%
\end{tabular}%
\end{table}%
%
%
The anticipation concerning the differences to CCS, formulated by \refnote{ccs_comparison}, find their substantiation in \reffig{performance_statistics}: on average and for all $\refvof$ under consideration, CCS requires one more truncation than the proposed algorithm. This can be attributed to the consecutive character of the spline interpolation, requiring, in general, two \textit{interior} truncations, i.e.\ at $\projvertposmin<\signdist^n,\signdist^{n+1}<\projvertposmax$. While \reffig{performance_statistics} only compares CCS for convex polyhedra, the manuscript of \citet{XXX_2020_ivof} also contains a parametrized set of non-convex polyhedra, for which CCS exhibits robust  performance and qualitatively equivalent results. %
%
%
In comparison, \citet{JCP_2016_anvc} only contains disaggregated data for convex polyhedra and \citet{JCP_2019_ncaa} contains no disaggregated data at all. For the tetrahedron, constituting the simplest possible polyhedron, the average number of truncations produced by CIBRAVE essentially coincides with those of the proposed algorithm. However, for the cuboid, the proposed method outperforms CIBRAVE by approximately \num{.5} truncations, on average. For non-convex polyhedra, no comparison could be made. Furthermore, recall that the sample set for the normals $\plicnormal$ in \citet{JCP_2016_anvc} is randomly computed, implying that degenerate cases (the plane $\plicplane$ is almost parallel to one of the faces $\vfface_k$, i.e.\ $\abs{\iprod{\plicnormal}{\vn_{\vfface,k}}}\approx1$), which are decisive for a meaningful examination, are not necessarily contained. %
The present method of rootfinding outperforms both CIBRAVE and CCS. While the absolute differences in terms of average truncations are small, taking into account the leverage effect of the reduced truncation complexity (and, thus, runtime) highlights the overall performance gain of the present algorithm. %
\item \refFig{newton_positioning_illustration} offers a qualitative illustration of a standard \textsc{Newton}-based method, similar to the algorithm of \citet{JCP_2019_apnm}. %
\begin{figure}[htbp]%
\null\hfill%
\subfigure[average number of truncations as a function of $\refvof$]{\label{fig:newton_positioning_illustration_01}
\includegraphics[page=1]{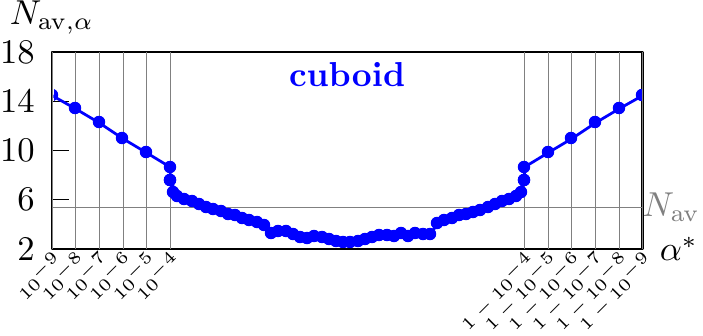}%
\hspace{1cm}%
\includegraphics[page=2]{average_iteration_newton}%
}
\hfill\null%
\\%
\null\hfill%
\subfigure[aggregated execution statistics]{\label{fig:newton_positioning_illustration_02}
\includegraphics{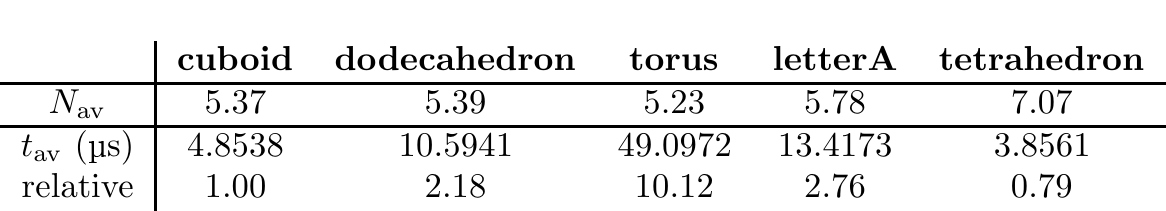}%
}%
\hfill\null%
\caption{Average number of truncations for \textsc{Newton}-based method of \protect\citet{JCP_2019_apnm}.}%
\label{fig:newton_positioning_illustration}%
\end{figure}%
As can be seen in the prototypical illustration of \reffig{newton_positioning_illustration_01}, besides the significant overall increase in the number of truncations by a factor of \num{4} to \num{5}, the \textsc{Newton}-based method also requires an increasing number of truncations for $\refvof\approx0$ and $\refvof\approx1$. This numerical artifact can be attributed to the diminishing derivatives $\polyvof^\prime$, resembling a well-known drawback of \textsc{Newton}-based methods. In a prototypical sense, this substantiates the superiority of spline-based methods over methods based on local first- or second-order approximation of $\polyvof\fof{\signdist}$. Among the different concepts of algorithmic incorporation, e.g., explicit bracketing (as employed by \citet{JCP_2016_anvc} in CIBRAVE) or consecutive interpolation (as employed by \citet{XXX_2020_ivof} in CCS), the impact on the average number of truncations is small in comparison to approximation-based methods. %
\end{enumerate}
\begin{figure}[htbp]
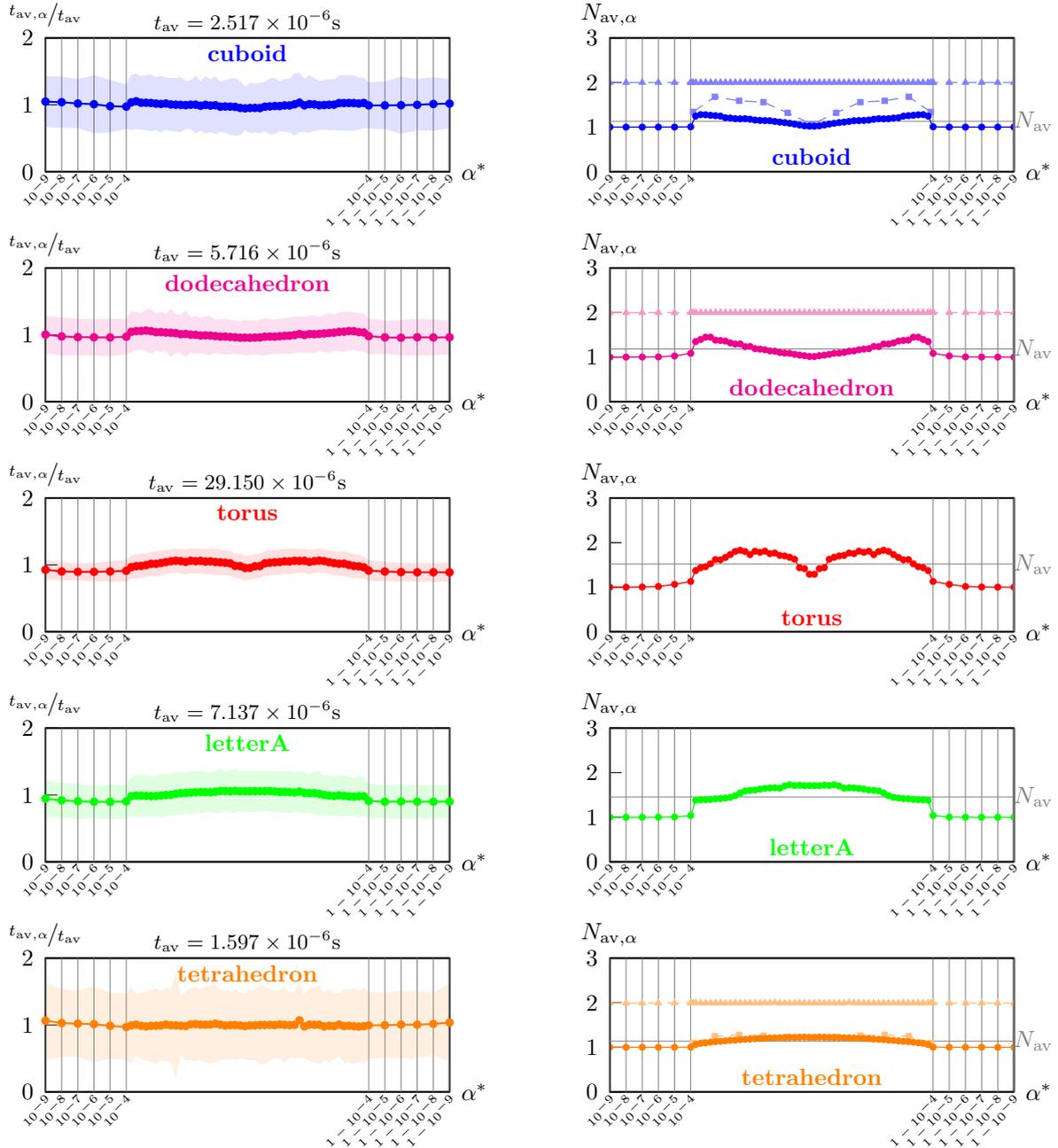

\multido{\i=1+1}{5}{%
\null\hfill%
\includegraphics[page=\i]{average_time}%
\hfill%
\includegraphics[page=\i]{average_iteration}%
\hfill\null%
\ifthenelse{\equal{\i}{5}}{}{\\}%
}%
\caption{Average positioning time in \si{\micro\second} and number of truncations for polyhedra from \reftab{polyhedron_types} with reconstruction tolerance $\epsilon=\num{e-12}$ obtained with proposed algorithm ($\bullet$), CCS ($\blacktriangle$) and CIBRAVE ($\blacksquare$).}%
\label{fig:performance_statistics}%
\end{figure}
%
%
\subsection{Influence of the normal orientation}\label{subsec:influence_normal}%
\begin{figure}[htbp]
\null\hfill%
\subfigure[\polyname{1}]{\includegraphics[page=1]{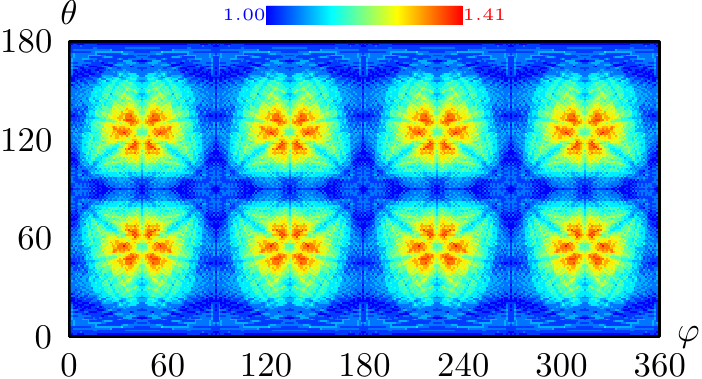}\label{fig:av_iter_spherical_01}}%
\hfill%
\subfigure[\polyname{2}]{\includegraphics[page=2]{statistics_normal}\label{fig:av_iter_spherical_02}}%
\hfill\null%
\\%
\null\hfill%
\subfigure[\polyname{4}]{\includegraphics[page=4]{statistics_normal}\label{fig:av_iter_spherical_03}}%
\hfill%
\subfigure[\polyname{5}]{\includegraphics[page=5]{statistics_normal}\label{fig:av_iter_spherical_04}}%
\hfill\null%
\caption{Average number of truncations over normal orientation in spherical coordinates.}%
\label{fig:av_iter_spherical}%
\end{figure}

\refFig{av_iter_spherical} associates the average number of truncations of the proposed method to the respective orientation of the normal $\plicnormal$, suggesting a parametrization in spherical coordinates. In general, the spatial distribution of the peaks in the average truncations resembles the geometry of the underlying polyhedron $\polyhedron*$, eventually inheriting some of its symmetries. \refFig{av_iter_spherical_01} vividly illustrates this for the cuboid: an orientation with $\theta\in\set{0,\nicefrac{\pi}{2},\pi}$ corresponds to a degeneration to two dimensions for arbitrary $\varphi$, implying that the associated volume fraction $\polyvof$ becomes quadratic. Since the proposed method locally applies a quadratic approximation, the computation of the sought position $\signdistref$ requires less truncations. On the contrary, for $\theta\not\in\set{0,\nicefrac{\pi}{2},\pi}$, the volume fraction is composed of non-degenerate cubic polynomials in general, imposing an increased computational effort to find $\signdistref$. %

%
%
\section{Summary \& Conlusions}\label{sec:summary}%
We have introduced an algorithm for the positioning of a PLIC plane in an arbitrary polyhedron. The recursive application of the \textsc{Gaussian} divergence theorem in its respectively appropriate form allows to obtain a highly efficient face-based computation of the volume of the truncated polyhedron, implying that no connectivity information has to be established at runtime. In combination with the precomputability of the involved coefficients, the face-based character renders the presented approach most suitable for parallel computations on unstructured meshes. Furthermore, we rigorously derive the mathematical foundations and examine the relevant limiting cases. %
The root-finding component relies on a combination of bracketing and local second-order approximation, allowing to compute the sought position to a user-defined precision ($\zerotol=\num{e-12}$). For both the main algorithm and its secondary components, we have provided flowcharts along with a comprehensive description. %
We have assessed our algorithm on a set of convex and non-convex polyhedra of genus zero and one (i.e., with a hole), which, to this extent, has not been done before. For all numerical instances, comprising a wide range of volume fractions $\num{e-9}\leq\refvof\leq1-\num{e-9}$ and normal orientations $\plicnormal\in\unitsphere$, one to two truncations are required on average. The variations with respect to $\refvof$ and $\plicnormal$ are small, indicating the robustness of the algorithm. In terms of the average number of truncations, being accountable for the major part of the computational effort, the proposed algorithm outperforms existing methods. Thus, we draw the following conclusions: %
\begin{enumerate}%
\item The recursive application of the \textsc{Gaussian} divergence theorem in appropriate form allows for an efficient computation of the volume of a truncated \textbf{arbitrary polyhedron}, along with its derivatives. This face-based problem decomposition allows to \textbf{avoid extracting topological connectivity}, which is advantageous in terms of both implementation complexity and computational effort. %
\item The local knowledge of the first to third derivatives \textbf{substantially accelerates the iterative root-finding}. For a comprehensive set of combinations of polyhedra, volume fractions and normal orientations, one to two truncations are required on average to position the PLIC plane. %
\end{enumerate}%
%
%
\renewcommand{\bibname}{References}%

\begin{center}%
\textsc{Acknowledgment}\\[2ex]%
The authors gratefully acknowledge financial support provided by the German Research Foundation (DFG) within the scope of \href{www.sfbtrr75.de}{SFB-TRR 75 (project number 84292822)}. Furthermore, the authors would like to thank \href{https://www.mma.tu-darmstadt.de/index/mitarbeiter_3/mitarbeiter_details_mma_43648.en.jsp}{Dr.-Ing.~Tomislav Mari\'c} for his advice concerning the design of numerical experiments. %
\end{center}%
\clearpage%
%
%
\begin{appendix}
\section{Supplementary information}\label{app:supplementary_information}%
\subsection{Rootfinding for a cubic polynomial}\label{app:rootfinding_spline}%
\begin{figure}[htbp]
\null\hfill%
\includegraphics{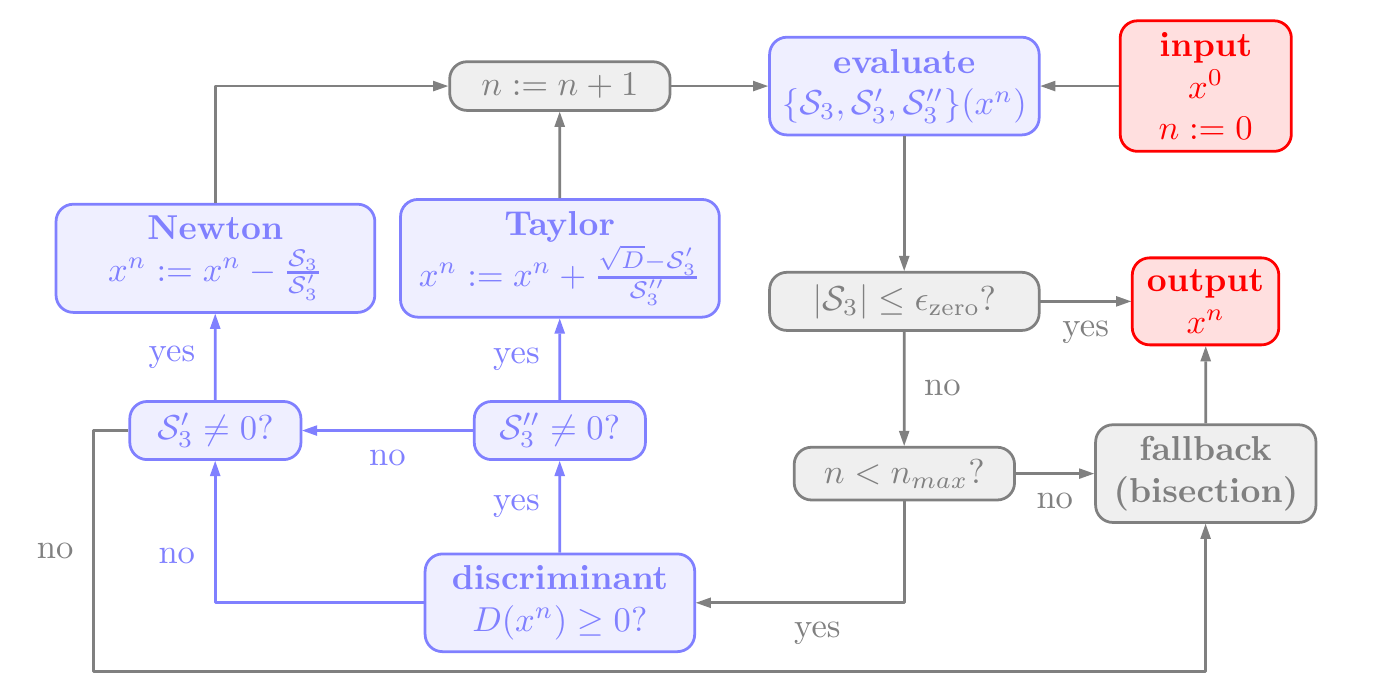}%
\hfill\null%
\caption{Flowchart of the cubic rootfinding algorithm.}%
\label{fig:rootfinding_cubic_flowchart}%
\end{figure}

The cubic spline tangentially interpolating the nodes $\set{\vp_1,\vp_2}$ with $\vp_i=\brackets[s]{x_i,f_i,f^\prime_i}$ and $\Delta x=x_j-x_i$ reads %
\begin{align}
\mathcal{S}_3\fof{x;\vp_i,\vp_j}=%
\frac{\Delta x\brackets{f^\prime_i+f^\prime_j}-2\brackets{f_j-f_i}}{\Delta x^3}\brackets{x-x_i}^3+%
\frac{3\brackets{f_j-f_i}-\Delta x\brackets{2f^\prime_i+f^\prime_j}}{\Delta x^2}\brackets{x-x_i}^2+%
f^\prime_i\brackets{x-x_i}+%
f_i.\label{eqn:cubic_spline_interpolation}%
\end{align}
In the non-gegenerate case (i.e., the cubic coefficient is non-zero), the root of \refeqn{cubic_spline_interpolation} is numerically computed using the algorithm given in \reffig{rootfinding_cubic_flowchart}, while in the (quadratic or linear) degenerate case the respective root is obtained analytically. Unless stated otherwise, the initial value is obtained by linear interpolation, i.e.\ $x^0\defeq x_i-\frac{f_i\Delta x}{f_j-f_i}$. %
\citet{JCP_2008_aagt} state that \glqq\textit{[...] explicit formulae for cubic roots generally consume less computational time than Newton--Raphson iterations}\grqq{}. However, they (i) do not specify the gain in terms of computational time and (ii) report seizing robustness for $\refvof<\num{e-6}$ and $\refvof>1-\num{e-6}$. Thus, we prefer to resort to the algorithm in \reffig{rootfinding_cubic_flowchart}. %
\subsection{Non-standard polyhedra}\label{app:nonstandard_polyhedron}
\begin{figure}[htbp]
\null\hfill%
\subfigure[Torus with relevant quantities.]{\label{fig:torus_illustration_quantities}\includegraphics{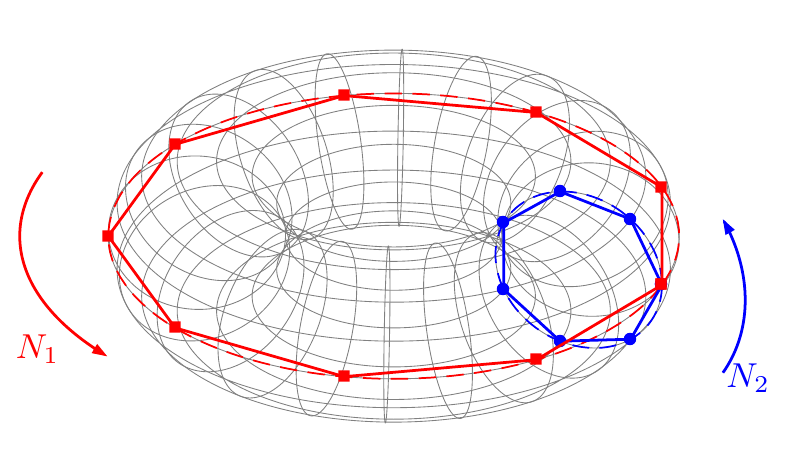}}%
\hfill%
\subfigure[Non-convex base faces of polyhedron \texttt{letterA}.]{\label{fig:letterA_illustration_vertices}\includegraphics{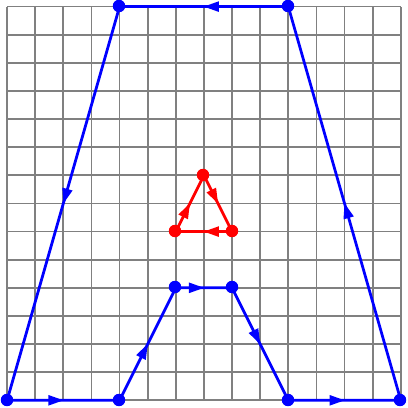}}%
\hfill\null%
\caption{Illustration of non-standard polyhedra.}%
\end{figure}

%
%
\paragraph{Torus.}%
The surface of a torus is parametrized by
\begin{align}
\vec{F}\fof{\varphi,\theta;R,\gamma}=R\brackets*[s]{%
\brackets{1+\gamma\cos\theta}\cos\varphi,%
\brackets{1+\gamma\cos\theta}\sin\varphi,%
\gamma\sin\theta
}\transpose,%
\end{align}
where $R$ is the major radius and $\gamma\in[0,1]$ is the minor radius ratio. The number of major (minor) nodes is denoted $N_1$ ($N_2$), such that the vertices of the torus are %
\begin{align}
\set*{\set*{\vec{F}\brackets*{\frac{2i\pi}{N_1},\frac{2j\pi}{N_2};R,\gamma}}_{j=1}^{N_2}}_{i=1}^{N_1},%
\end{align}%
where, throughout his work, we employ $\gamma=\nicefrac{1}{2}$, $R=1$, $N_1=9$ and $N_2=7$; see figures~\ref{fig:illustration_vof_02} and \ref{fig:torus_illustration_quantities} for an illustration. %
%
%
\paragraph{Letter A.}%
The vertices of the letter A are $$\textcolor{blue}{\nicefrac{1}{14}\set{(0,0,d),(4,0,d),(6,4,d),(8,4,d),(10,0,d),(14,0,d),(10,14,d),(8,14,d)}}$$ and $$\textcolor{red}{\nicefrac{1}{14}\set{(6,6,d),(8,6,d),(7,8,d)}},$$%
with $d\in\set{0,5}$ corresponds to the front and back face, respectively. \refFig{letterA_illustration_vertices} provides an illustration. The hole is created by an inverted arrangement of the blue and red set of vertices, resulting in two coplanar faces with opposing normals; cf.~\reffig{polygon_hole_definition}. %
%
%
\end{appendix}%
\end{document}